\documentclass[sigconf]{acmart}

\copyrightyear{2020}
\acmYear{2020}
\setcopyright{iw3c2w3}
\acmConference[WWW '20]{Proceedings of The Web Conference 2020}{April 20--24, 2020}{Taipei, Taiwan}
\acmBooktitle{Proceedings of The Web Conference 2020 (WWW '20), April 20--24, 2020, Taipei, Taiwan}
\acmPrice{}
\acmDOI{10.1145/3366423.3380121}
\acmISBN{978-1-4503-7023-3/20/04}

\usepackage[utf8]{inputenc} % allow utf-8 input
\usepackage[T1]{fontenc}    % use 8-bit T1 fonts
\usepackage{hyperref}       % hyperlinks
\usepackage{url}            % simple URL typesetting
\usepackage{booktabs}       % professional-quality tables
\usepackage{amsfonts}       % blackboard math symbols
\usepackage{nicefrac}       % compact symbols for 1/2, etc.
\usepackage{microtype}      % microtypography
\usepackage{amsmath}
\usepackage{amssymb}
\usepackage{mathtools}
\usepackage{natbib}
\usepackage{graphicx}
\usepackage{url}
\usepackage{xspace}
\usepackage{float}
\usepackage{tikz}
\usetikzlibrary{calc,positioning}
\usepackage{varwidth}
\usepackage{caption}
\usepackage{enumitem}
\usepackage{pgfplots}
\pgfplotsset{compat=newest}
\usepgfplotslibrary{groupplots}
\usepgfplotslibrary{dateplot}

\makeatletter
\newcommand\footnoteref[1]{\protected@xdef\@thefnmark{\ref{#1}}\@footnotemark}
\makeatother

\usepackage{amsthm}
\usepackage[ruled, vlined, nofillcomment, linesnumbered]{algorithm2e}
\usepackage[hide]{chato-notes}

\newtheorem*{problem*}{Problem}
\newtheorem{theorem}{Theorem}
\newtheorem{proposition}{Proposition}

\newtheorem{lemma}{Lemma}

\newtheorem{claim}{Claim}

\DeclareMathOperator*{\argmin}{arg\,min}

\newcommand{\set}[1]{\left\{#1\right\}}

\newcommand{\abs}[1]{{\left|#1\right|}}

\newcommand{\enlst}[2]{{#1} ,\ldots , {#2}}

\newcommand{\real}{\mathbb{R}}

\newcommand{\define}{\leftarrow}

\newcommand{\inprodD}[2]{\langle\,#1,#2\rangle_{\dmat}\xspace}

\let\emptyset\varnothing
\newcommand{\mrange}{\mathcal{R}\xspace}
\newcommand{\mnull}{\mathcal{N}\xspace}
\newcommand{\trace}{\mathit{tr}\xspace}

\newcommand{\expplus}{\ensuremath{\text{+}}\xspace}
\newcommand{\expminus}{\ensuremath{\text{--}}\xspace}

\newcommand{\graph}{\ensuremath{G}\xspace}

\newcommand{\nodes}{\ensuremath{V}\xspace}
\newcommand{\edges}{\ensuremath{E}\xspace}

\newcommand{\posedges}{\ensuremath{\edges^{\expplus}}\xspace}
\newcommand{\negedges}{\ensuremath{\edges^{\expminus}}\xspace}
\newcommand{\graphdef}{\ensuremath{(\nodes, \posedges, \negedges)}\xspace}

\newcommand{\bigO}{\ensuremath{\mathcal{O}}\xspace}

\newcommand{\adjm}{\ensuremath{A}\xspace}
\newcommand{\posadjm}{\ensuremath{\adjm^{\expplus}}\xspace}
\newcommand{\negadjm}{\ensuremath{\adjm^{\expminus}}\xspace}

\newcommand{\degree}{\ensuremath{\mathit{deg}}\xspace}

\newcommand{\posdegree}{\ensuremath{\degree^{\expplus}}\xspace}

\newcommand{\cut}{\ensuremath{\mathit{cut}}\xspace}
\newcommand{\poscut}{\ensuremath{\mathit{cut}^{\expplus}}\xspace}
\newcommand{\negcut}{\ensuremath{\mathit{cut}^{\expminus}}\xspace}
\newcommand{\posin}{\ensuremath{\mathit{in}^{\expplus}}\xspace}
\newcommand{\negin}{\ensuremath{\mathit{in}^{\expminus}}\xspace}

\newcommand{\vol}{\ensuremath{\mathit{vol}}}
\newcommand{\posvol}{\ensuremath{\mathit{vol}^{\expplus}}}
\newcommand{\negvol}{\ensuremath{\mathit{vol}^{\expminus}}}

\newcommand{\dmat}{\ensuremath{D}\xspace}
\newcommand{\Dneghalf}{\ensuremath{\dmat^{-\frac{1}{2}}}\xspace}
\newcommand{\Dhalf}{\ensuremath{\dmat^{\frac{1}{2}}}\xspace}
\newcommand{\lap}{\ensuremath{L}\xspace}
\newcommand{\nlap}{\ensuremath{\mathcal{L}}\xspace}
\newcommand{\eigval}{\ensuremath{\lambda_1}\xspace}
\newcommand{\evec}{\ensuremath{\mathbf{v}_1}\xspace}
\newcommand{\evecT}{\ensuremath{\mathbf{v}_1^T}\xspace}
\newcommand{\evectwo}{\ensuremath{\mathbf{v}_2}\xspace}

\newcommand{\sbr}{\ensuremath{\beta}\xspace}
\newcommand{\cheegerc}{\ensuremath{h}\xspace}
\newcommand{\vx}{\ensuremath{\mathbf{x}}\xspace}
\newcommand{\vxopt}{\ensuremath{\vx^{*}}\xspace}
\newcommand{\vs}{\ensuremath{\mathbf{s}}\xspace}
\newcommand{\vy}{\ensuremath{\mathbf{y}}\xspace}
\newcommand{\Sone}{\ensuremath{C_1}\xspace}
\newcommand{\Stwo}{\ensuremath{C_2}\xspace}
\newcommand{\Sonetrue}{\ensuremath{C_1^{*}}\xspace}
\newcommand{\Stwotrue}{\ensuremath{C_2^{*}}\xspace}
\newcommand{\Sione}{\ensuremath{C_{i,1}}\xspace}
\newcommand{\Sitwo}{\ensuremath{C_{i,2}}\xspace}
\newcommand{\Srest}{\ensuremath{\nodes\setminus(\Sone \cup \Stwo)}\xspace}
\newcommand{\rquot}{\ensuremath{\mathcal{R}_{\nlap}}\xspace}
\newcommand{\vxab}{\ensuremath{\vs_{\Sone, \Stwo}}\xspace}
\newcommand{\vone}{\ensuremath{1_{\Sone}}\xspace}
\newcommand{\vtwo}{\ensuremath{1_{\Stwo}}\xspace}
\newcommand{\vA}{\ensuremath{1_A}\xspace}
\newcommand{\vB}{\ensuremath{1_B}\xspace}
\newcommand{\vsAB}{\ensuremath{\vs_{A, B}}\xspace}
\newcommand{\vu}{\ensuremath{1_{\seedp}}\xspace}
\newcommand{\vv}{\ensuremath{1_{\seedn}}\xspace}
\newcommand{\vsuv}{\ensuremath{\vs_{\seedp, \seedn}}\xspace}

\newcommand{\geneq}{\succeq}
\newcommand{\mprod}{\circ}

\newcommand{\pinv}[1]{{(#1)}^{\dagger}}
\newcommand{\dsds}{(D \vs) (D \vs)^T}
\newcommand{\Xopt}{X^{*}}
\newcommand{\xopt}{\vx^{*}}
\newcommand{\alphaopt}{\alpha^{*}}
\newcommand{\betaopt}{\beta^{*}}

\definecolor{myred}{rgb}{0.98, 0.60, 0.6}
\definecolor{mygreen}{rgb}{0.69, 0.87, 0.54}
\definecolor{myblue}{rgb}{0.65, 0.80, 0.89}

\newcommand{\localpol}{\textsc{Local\-Polar($\graph, \vs, \kappa$)}\xspace}
\newcommand{\localpoldiscrete}{\textsc{Local\-Polar\-Discrete($\graph, \seedp, \seedn, k$)}\xspace}
\newcommand{\localpoldiscreteshort}{\textsc{Local\-Polar-Discrete}\xspace}
\newcommand{\localpols}{\textsc{Local\-Polar}\xspace}

\newcommand{\localpolk}{\textsc{Local\-Polar($\graph, \vsuv, \sqrt{1/k}$)}\xspace}
\newcommand{\sdpp}{\textsc{SDP$_p$($\graph, \vs, \kappa$)}\xspace}

\newcommand{\sdpps}{\textsc{SDP$_p$}\xspace}
\newcommand{\sdpd}{\textsc{SDP$_d$($\graph, \vs, \kappa$)}\xspace}

\newcommand{\tol}{\ensuremath{\epsilon}\xspace}

\newcommand{\focg}{\textsc{\large focg}\xspace}
\newcommand{\ours}{\textsc{Polar\-Seeds}\xspace}

\newcommand{\dsword}{\emph{Word}\xspace}
\newcommand{\dsbc}{\emph{Bitcoin}\xspace}
\newcommand{\dsref}{\emph{Referendum}\xspace}
\newcommand{\dssd}{\emph{Slashdot}\xspace}
\newcommand{\dsep}{\emph{Epinions}\xspace}
\newcommand{\dswiki}{\emph{WikiConflict}\xspace}
\newcommand{\dswikilarge}{\emph{WikiConflict-1M}\xspace}

\newcommand{\seeds}{\ensuremath{S}\xspace}
\newcommand{\seedp}{\ensuremath{S_1}\xspace}
\newcommand{\seedn}{\ensuremath{S_2}\xspace}

\newcommand{\commcount}{\ensuremath{p}\xspace}

\newcommand{\noise}{\ensuremath{\eta}\xspace}
\newcommand{\seedcount}{\ensuremath{q}\xspace}

\newcommand{\ap}{\ensuremath{\text{\sc\large ap}}\xspace}

\newcommand{\ham}{\ensuremath{\text{\sc\large ham}}\xspace}
\newcommand{\coh}{\ensuremath{\text{Cohesion}}\xspace}
\newcommand{\opp}{\ensuremath{\text{Opposition}}\xspace}
\newcommand{\pc}{\ensuremath{\text{Polarity}}\xspace}

\newcommand{\Ap}{\ensuremath{\posadjm}\xspace}
\newcommand{\An}{\ensuremath{\negadjm}\xspace}
\newcommand{\Lp}{\ensuremath{L^{+}}\xspace}
\newcommand{\Ln}{\ensuremath{L^{-}}\xspace}

\newcommand{\domain}{\ensuremath{\text{unique}}\xspace}
\newcommand{\nvx}{\ensuremath{-\vx}\xspace}
\newcommand{\avx}{\ensuremath{\abs{\vx}}\xspace}

\newcommand{\Cavx}{\ensuremath{C_{\avx}}\xspace}
\newcommand{\Cvx}{\ensuremath{C_{\vx}}\xspace}
\newcommand{\Cnvx}{\ensuremath{C_{\nvx}}\xspace}

\newcommand{\spara}[1]{\smallskip\noindent{\bf{#1}}}

\newcommand{\squishlist}{
 \begin{list}{$\bullet$}
  {  \setlength{\itemsep}{0pt}
     \setlength{\parsep}{3pt}
     \setlength{\topsep}{3pt}
     \setlength{\partopsep}{0pt}
     \setlength{\leftmargin}{2em}
     \setlength{\labelwidth}{1.5em}
     \setlength{\labelsep}{0.5em}
} }
\newcommand{\squishlisttight}{
 \begin{list}{$\bullet$}
  { \setlength{\itemsep}{0pt}
    \setlength{\parsep}{0pt}
    \setlength{\topsep}{0pt}
    \setlength{\partopsep}{0pt}
    \setlength{\leftmargin}{2em}
    \setlength{\labelwidth}{1.5em}
    \setlength{\labelsep}{0.5em}
} }

\newcommand{\squishdesc}{
 \begin{list}{}
  {  \setlength{\itemsep}{0pt}
     \setlength{\parsep}{3pt}
     \setlength{\topsep}{3pt}
     \setlength{\partopsep}{0pt}
     \setlength{\leftmargin}{1em}
     \setlength{\labelwidth}{1.5em}
     \setlength{\labelsep}{0.5em}
} }

\newcommand{\squishend}{
  \end{list}
}

\graphicspath{{./figs/}}

\AtBeginDocument{%
  \providecommand\BibTeX{{%
    \normalfont B\kern-0.5em{\scshape i\kern-0.25em b}\kern-0.8em\TeX}}}

\setcopyright{acmcopyright}

\begin{document}

\title[Local polarized communities in signed graphs]{Searching for polarization in signed graphs: \\a local spectral approach}


\author{Han Xiao}
\affiliation{%
  \institution{Aalto University}
  \streetaddress{}
  \city{Espoo}
  \state{Finland}
  \postcode{}
}
\email{han.xiao@aalto.fi}

\author{Bruno Ordozgoiti}
\affiliation{%
  \institution{Aalto University}
  \streetaddress{}
  \city{Espoo}
  \state{Finland}
  \postcode{}
}
\email{bruno.ordozgoiti@aalto.fi}

\author{Aristides Gionis}
\authornote{This work was done while the author was with Aalto University.}
\affiliation{%
  \institution{KTH Royal Institute of Technology}
  \streetaddress{}
  \city{Stockholm}
  \state{Sweden}
  \postcode{}
}
\email{argioni@kth.se}


\begin{abstract}
Signed graphs have been used to model interactions in social networks, 
which can be either positive (friendly) or negative (antagonistic).
The model has been used to study polarization and other related phenomena in social networks,  
which can be harmful to the process of democratic deliberation in our society. 
An interesting and challenging task in this application domain is to detect 
polarized communities in signed graphs.
A number of different methods have been proposed for this task. 
However, existing approaches aim at finding \emph{globally} optimal solutions. 
Instead, in this paper we are interested in finding polarized communities that are \emph{related to a small set of seed nodes provided as input}.
Seed nodes may consist of two sets,
which constitute the two sides of a polarized structure.

In this paper we formulate the problem of finding local polarized communities in signed graphs
as a \emph{locally-biased eigen-problem}. 
By viewing
the eigenvector associated with the \textit{smallest} eigenvalue of the Laplacian matrix
as the solution of a constrained optimization problem, 
we are able to incorporate the local information as an additional constraint. 
In addition, we show that the locally-biased vector can be used to find communities with approximation guarantee 
with respect to a local analogue of the Cheeger constant on signed graphs. 
By exploiting the sparsity in the input graph, an indicator-vector for the polarized communities 
can be found in time linear to the graph size. 
  
Our experiments on real-world networks validate the proposed algorithm and demonstrate its usefulness in finding local structures in this semi-supervised manner.

% \cnote{
% Aris: I think it is important to explicitly highlight the differences between our derivation and Mahoney's. For instance, a key difference is that we don't have $x^Tv_1=0$
% }
% \cnote{
% Han: can we introduce polarization a bit? like its societal-level prevalence and harm.} 
% \cnote{Aris: I tried to motivate the problem in the application of polarization.}
\end{abstract}

\maketitle

\section{Introduction}
The issue of polarization in social media is becoming of increasing interest, due to its impact on the health of public discourse and the integrity of democratic processes. Detecting polarized structures in social networks is therefore a well-motivated problem. One way to accomplish this task is to employ graph clustering \cite{schaeffer2007graph} or community-detection methods \cite{fortunato2010community}. These techniques help us locate dense structures in networks, which might correspond to so-called echo chambers~\cite{garimella2018quantifying}.

Spectral methods are commonly used for graph clustering. By inspecting the eigenvectors of the graph Laplacian it is possible to find partitions with quality guarantees \cite{chungfour}, often formulated in terms of conductance---a well-known measure of graph-partition quality. A shortcoming of these techniques is that they are inherently global, and therefore not useful for finding local solutions, which could be relevant to the interests of a user. To overcome this limitation, Mahoney et al.\ propose a locally-biased community-extraction formulation \cite{mahoney2012local}. The idea is to find a solution that is correlated with a given input vector. The approach enables us to detect dense communities that are close to a subgraph of interest. Previous approaches for similar tasks employ PageRank vectors~\cite{andersen2006local, bian2018multi, gleich2012vertex, kloster2014heat}. 

To encode the attitude or sentiment of interactions between individuals in a social network, we can annotate every edge in the corresponding graph with a positive or negative sign, so as to signify whether the interaction is friendly or hostile. A graph featuring such edge annotations is known as a \textit{signed graph}. The analysis of signed graphs, in particular for clustering and community detection, has received considerable attention in recent years~\cite{chu2016finding, kunegis2010spectral, leskovec2010signed, tang2016survey}. 
A number of different methods have been proposed for finding polarized communities in signed graphs~\cite{chu2016finding, lo2013mining}, however, existing methods aim at finding \emph{globally} optimal solutions. 
To our knowledge, this is the first paper that addresses the problem of finding
local polarized communities in signed graphs.

% \cnote{Aris: Verify that the claim above is correct}
% \cnote{Han: to my knowledge, yes}
% \cnote{Todo: describe 3 applications: knowledge discovery in general, reduce polarization and finding overlapping community}
% \cnote{Aris: I made some small edits and moved the paragraph before contributions.}

\spara{Motivation.}
Finding polarized groups related to a set of seed nodes has several important applications.
As an example, consider a scenario in which a polarized debate emerges around a topic discussed in an online social-media platform (e.g., Facebook).
Assume further that a small number of high-profile users are known to be engaged in hostile interactions.
Our algorithm could then determine the remaining members of the two rival groups, 
using the set of known users as a seed.
Finding the two polarized communities can be useful for computational social scientists who are interested in understanding 
the structure of such communities and the mechanics of their discussions~\cite{vegetti2019political}.
In addition, finding the polarized communities can be of interest to the social-media platform, 
if they wish to reduce the degree of polarization by recommending a thought-provoking article to the 
users involved in the argument.

\begin {figure}[t]
  \centering
  \begin{tabular}{cc}
    \includegraphics[width=.22\textwidth]{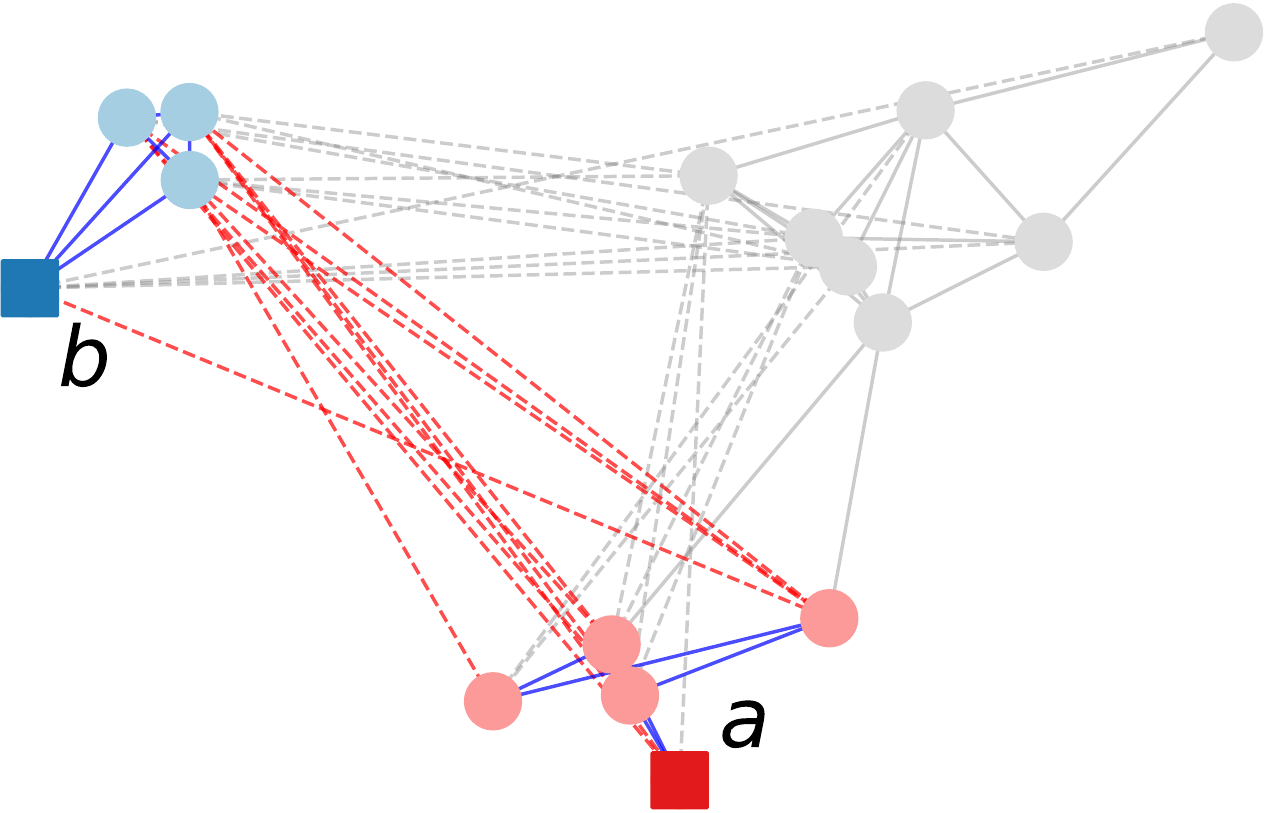} & \includegraphics[width=.22\textwidth]{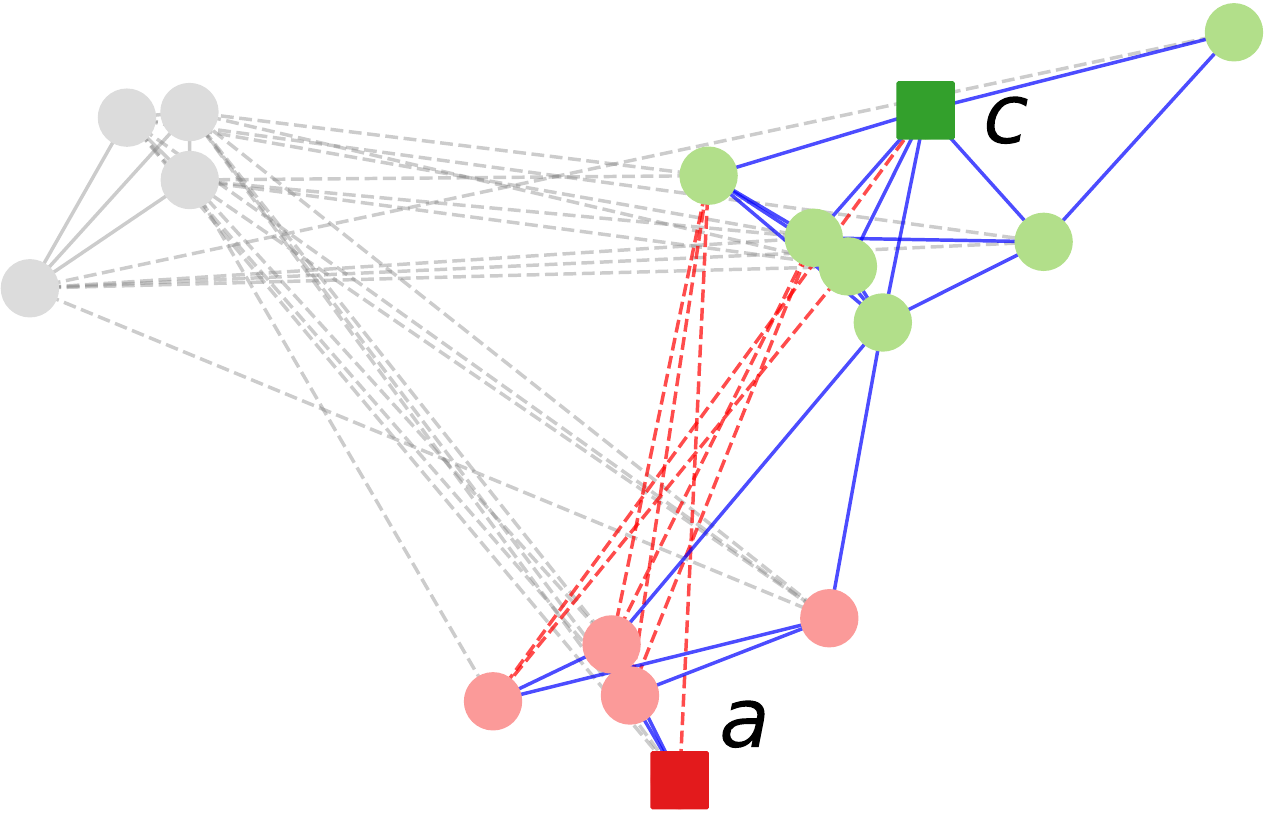} \\
    (a) query $\seedp=\set{a}$ and $\seedn=\set{b}$ & (b) query $\seedp=\set{a}$ and $\seedn=\set{c}$ 
  \end{tabular}
  \caption{A motivating example of local polarization search on the \textit{New Guinea Highland Tribes} graph~\cite{read1954cultures}:
    individual tribes are shown as vertices, with friendly relations shown as solid line and enemy relations as dashed line.
    Given a pair of node sets $(\seedp, \seedn)$ as query,
    we are interested in finding two subgraphs $\Sone$ and $\Stwo$ such that:
    ($i$) $\Sone$ and $\Stwo$ are antagonistic to each other
    ($ii$) $\Sone$ and $\Stwo$ are friendly inside themselves
    and ($iii$) nodes in $\Sone$ and $\Stwo$ are related to  $\seedp$ and $\seedn$. 
    In the figure, we show two querying scenarios. 
    (a) querying $\seedp=\set{a}$ and $\seedn=\set{b}$ produces $\Sone$ and $\Stwo$ as the red and blue nodes respectively.
    (b) querying $\seedp=\set{a}$ and $\seedn=\set{c}$ produces $\Sone$ and $\Stwo$ as the red and green nodes.
    Note that the two pairs of subgraphs are overlapping in red-colored nodes.
  }
  \label{fig:motivation}
\end {figure}

\spara{Contributions.} In this paper we propose a local spectral method to find polarized communities in signed graphs. In particular, given a set of seed nodes, our method finds two antagonistic groups close to the seed nodes. 
We illustrate such a scenario in Figure~\ref{fig:motivation}. 
From a technical point of view, 
we extend the framework of Mahoney et al. \cite{mahoney2012local} to deal with signed graphs, 
whose spectral properties exhibit key differences with respect to their unsigned counterpart.

By formulating the problem as a convex semi-definite program
and utilizing duality theory, we are able to find the closed-form of the optimal solution. 
Moreover, we combine the resulting method with a Cheeger-type inequality for signed graphs to show that our algorithm enjoys provable approximation guarantees. In particular, given two sets of seed nodes $\seedp, \seedn$ and a correlation parameter $\kappa$ our method finds a solution whose \textit{signed bipartiteness ratio}, a signed analogue of conductance, is $\bigO(\sqrt{\cheegerc(\seedp,\seedn,k)})$, where $\cheegerc(\seedp,\seedn,k)$ is the optimum with respect to \seedp and \seedn, and $k = 1/\kappa^2$.
% $\kappa=\sqrt{1/k}$.

\iffalse
The application domain extends to human vocabulary networks. 
In our case studies (Section~\ref{subsec:casestudy}), we are able to find sets of synonyms that have opposite meaning to each other. 
\fi 

\subsection{Related work}

\spara{Local community detection.}
The problem of finding local communities with respect to seed nodes has been studied extensively. 
In general, given a set of seed nodes $\seeds$ the task is to find a subgraph that contains $\seeds$ and optimizes some quality measure. 
Various measures have been proposed, such as
conductance~\cite{andersen2006local}, 
$k$-core~\cite{sozio2010community},
$k$-truss~\cite{huang2014querying},
and quasi-cliques~\cite{cui2014local}.
One line of effort utilizes PageRank methods~\cite{andersen2006local, gleich2012vertex, kloster2014heat, bian2018multi}, which enjoy provable performance guarantee~\cite{andersen2006local} and are effective in practice~\cite{kloumann2014community}. 
Mahoney et al.~\cite{mahoney2012local} combine seed information with spectral partitioning and obtain guarantees with respect to conductance.

% \cnote{Han: elaborated on the difference between our approach and Mahoney's paper}
% \cnote{Aris: I edited the section. Please check again.}

Our approach is inspired by the work of Mahoney et al.~\cite{mahoney2012local}, 
and our paper extends their results to signed graphs.
Yet, to make the extension possible, several technical results are introduced.
First, given the spectral properties of signed graphs, we work with $\evec$, the eigenvector of the \textit{smallest} eigenvalue, while Mahoney’s work use $\evectwo$, the eigenvector of the \textit{second smallest} eigenvalue.
This key difference leads to a different problem, as the solution is no longer orthogonal to the first vector.
In addition, different intermediate results and proofs are required. 
Second, we rely on recent results on signed graphs (i.e., Proposition 1 in Section~\ref{sec:proof}), not present in the literature on unsigned graphs to the best of our knowledge.
Third, efficient rounding of fractional solutions is more complicated on signed graphs than on unsigned graphs, 
and new techniques are necessary.

\spara{Spectral graph theory on signed graphs.}
Spectral methods have been studied extensively also for signed graphs.
Kunegis et al.~\cite{kunegis2010spectral} study normalized cuts, while Chiang et al.~\cite{chiang2012scalable} study $k$-way partitioning.
Mercado et al.~\cite{mercado2016clustering} propose the use of the geometric mean of the Laplacians of the positive and negative parts of the graph to correctly cluster graphs under the stochastic block model.
Note that all these methods are global, while our method is local. 

The spectrum of the signed Laplacian reveals interesting properties about the corresponding graph.
The smallest eigenvalue bounds the \emph{frustration number} (resp.\ \emph{index})~\cite{belardo2014balancedness,li2009note}, which is the minimum number of nodes (resp.\ edges) to remove to make graph \textit{balanced}.
% i.e. perfectly partitionable with respect to edge signs.
Notably, Atay et al.~\cite{atay2020cheeger} shows that the smallest eigenvalue of the normalized Laplacian is closely related to \textit{signed bipartiteness ratio}, which can be seen as a signed analogue of conductance. 
% Together, they form the keys of signed Cheeger's inequality. 

\spara{Antagonistic communities.}
Finding antagonistic communities in signed graphs~\cite{chu2016finding,gao2016detecting,lo2013mining,bonchi2019discovering} has attracted attention in recent years.
Existing works differ in the definition of an antagonistic community. 
% We highlight the fact that the previously cited 
Most works focus on \textit{enumerating} antagonistic communities at the global level, 
and implicitly reduce information redundancy by avoiding overlapping communities.
In contrast, 
we focus on \textit{locating} antagonistic communities given seed nodes.
It should be noted that our setting is one approach to deal with overlapping structures, 
e.g., as depicted in Figure~\ref{fig:motivation}. 
%  and meanwhile, we embrace overlapping structures, which is prevalent in real world graphs~\cite{yang2014overlapping}. 
% \cnote{chu2016finding can find overlapping bands, so maybe we should walk back on this claim.}
Furthermore, our approach yields approximation guarantees in terms of our definition of antagonism,
while previous methods employ heuristics without quality guarantees. 

\iffalse
  \spara{Contributions} We summarize our contributions below:
  \begin{itemize}
  \item We propose a local spectral method to find polarized communities in signed networks.
  \item We formulate the problem as a convex semi-definite program use duality theory to find an optimal solution
  \item We show the algorithm achieves a signed bipartiteness ratio of $\bigO(\sqrt{\cheegerc(u,v,k)})$, where $\cheegerc(u,v,k)$ is the optimum.
  \item We demonstrate the effectiveness and the efficiency of our approach in a variety of experiments on real and synthetic data, where it shows superior performance than comparable methods from the literature.
  \end{itemize}
\fi

\section{Problem formulation}

We represent column vectors by boldface lower-case letters, e.g., $\vx, \vy$, 
and matrices by upper-case letters, e.g., $\adjm$, $\dmat$.
We write $\vx_i$ to denote the $i$-th entry in vector $\vx$ and $A_{i, j}$ to denote the $j$-th entry in the $i$-th row of matrix $A$.
Our notation is shown in Table~\ref{tbl:symbol}.

\begin{table}[t]
  \centering
  \begin{small}
  \caption{\label{tbl:symbol}Notation}
  \vspace{-4mm}
%  \resizebox{1\columnwidth}{!}{
    \begin{tabular}{ll}
      \toprule
      \textbf{Symbol} &   \textbf{Meaning} \\
      \midrule
      $\posedges, \negedges$ & positive/negative edges \\
      $\posadjm, \negadjm, \adjm$ & positive/negative/signed adjacency matrix \\
      $\dmat$ & degree matrix \\
      $\vol(C)$ & volume of node set $C$ \\
      \midrule      
      $\lap, \nlap$ & unnormalized/normalized signed Laplacian matrix \\
      $\eigval, \evec$ & smallest eigenvalue of $\nlap$ and its eigenvector \\
      \midrule
      $\Sone, \Stwo$ & two bands of a polarized community \\
      $\seedp, \seedn$ & seed sets corresponding to the two bands \\
      $\sbr(\Sone, \Stwo)$ & signed bipartiteness ratio of $\Sone, \Stwo$\\
      $k$ &  volume parameter in \localpoldiscreteshort\\
      $\vs$ & seed vector in \localpols \\
      $\kappa$ & correlation parameter in \localpols \\
      \midrule
      % $\cheegerc(\graph)$ & signed Cheeger's constant of $\graph$ \\
      $\cheegerc(\seedp, \seedn, k)$ & local signed Cheeger's constant w.r.t $\seedp, \seedn$ and $k$\\
      $\lambda(\vs, \kappa)$ & locally-biased smallest eigenvalue w.r.t.\ $\vs$ and $\kappa$\\
      \bottomrule
    \end{tabular}
  \end{small}
%  }
\end{table}

\spara{Signed graphs.}
A \textit{signed graph} is defined as an undirected graph $\graph=\graphdef$,
where $\nodes=\set{1, \ldots, n}$ is a set of nodes, 
\posedges a set of positive edges, and 
\negedges a set of negative edges.
The set of all edges is $\edges=\posedges \cup \negedges$. 
The adjacency matrix of the positive edges of the graph is denoted by \posadjm.
The entry $\posadjm_{i, j}$ is defined to be 1 (or another positive value if the graph is weighted) 
if $(i, j)\in\posedges$, and 0 otherwise. 
The adjacency matrix of the negative edges of the graph is denoted by \negadjm and is defined  a similar way.
We combine \posadjm and \negadjm into a single term,
the \textit{signed adjacency matrix} $\adjm=\posadjm - \negadjm$.
We focus on undirected graphs, so \posadjm, \negadjm, and \adjm are all symmetric. 

\spara{Spectral graph theory on signed graphs.}
% We next review some relevant facts from spectral graph theory on signed graphs.
Given a signed graph $\graph=(\nodes, \posedges, \negedges)$ we define the following matrices: 
the \textit{degree matrix} $\dmat$ is a diagonal matrix with $\dmat_{i, i}=\degree(i)=\sum_{j=1}^n |\adjm_{i,j}|$;
the \textit{signed Laplacian matrix} is defined as $\lap=\dmat-\adjm$.
Normalizing $\lap$ gives the \textit{signed normalized Laplacian}
$\nlap=\Dneghalf \lap \Dneghalf = I-\Dneghalf A \Dneghalf$.%
\footnote{We ignore the prefix ``signed'' when it is clear from the context.}
% We denote the smallest eigenvalue of $\nlap$ as \eigval
% and the corresponding eigenvector as $\evec$.

Given a vector $\vx \in \real^{n}$ and associated vector $\vy=\Dhalf \vx$, 
we define the \textit{Rayleigh quotient} $\rquot(\vy)$:
\begin{equation}
\rquot(\vy) =  \frac{\vy^T \nlap \vy}{\vy^T \vy} = \frac{\vx^T \lap \vx}{\vx^T \dmat \vx}.
\end{equation}
Minimizing $\rquot(\vy)$ yields the smallest eigenvalue $\eigval$ of $\nlap$ and its associated eigenvector $\evec$.
If $\graph$ is connected, a small value of $\eigval$ indicates a more ``balanced'' graph, 
where a graph $\graph$ is \textit{perfectly balanced} if its nodes can be partitioned into two sets $\Sone$ and $\Stwo$,
such that all edges within $\Sone$ and $\Stwo$ are positive and all edges across are negative.
If $\graph$ is perfectly balanced, $\eigval$ is zero. Otherwise, we say it is \textit{unbalanced}.
% In the general case, 
The eigenvector $\evec$ gives hints on a node partitioning to minimize $\rquot$.
% e.g., $i \in \Sone$ if $\evec(i) > 0$ and $i \in \Stwo$ otherwise. 

\spara{Measuring polarization.}
We define a \textit{polarized community} $(\Sone, \Stwo)$ 
as two disjoint sets of nodes $\Sone, \Stwo \subseteq \nodes$, such that
% ($i$) there are relatively many edges within $\Sone$ and within $\Stwo$ and they are mostly positive; 
% ($ii$) there are relatively many edges between $\Sone$ and $\Stwo$ and they are mostly negative;
($i$) there are relatively few (resp. many) negative (resp. positive) edges within $\Sone$ and within $\Stwo$; 
($ii$) there are relatively few (resp. many) positive (resp. negative) edges across $\Sone$ and $\Stwo$;
and
($iii$) there are relatively few edges (of either sign) from $\Sone$ and $\Stwo$ to the rest of the graph.
We refer to $\Sone$ and $\Stwo$ as the two {\em bands} of the community.

% \cnote{We refer to $\Sone$ and $\Stwo$ as the two {\em bands} of the community.}

Before formalizing the degree of polarization of a community, we introduce some additional notation. 
Given a set of edges $F$ and two node sets $X, Y \subseteq \nodes$,
we consider the edges of $F(X, Y)$ that have one endpoint in $X$ and one in $Y$, i.e., 
$F(X, Y) = \set{(u, v) \in F \mid u \in X, v \in Y}$. 
We define $F(X) = F(X, X)$.
For example, $\posedges(X)$ is the set of positive edges having both endpoints in $X$.
The \textit{volume} of a node set $S$ is defined as 
$\vol(S) = \sum_{i \in S} \degree(i)$. 

% \cnote{Bruno: Removed "By setting $F=\posedges, \negedges \text{ or } \edges$, we can refer to 
% positive, negative, or unsigned edges between different node sets." I paste it here in case you want to easily revert.}

We also represent a community $(\Sone,\Stwo)$ in vector form. 
Given bands \Sone and \Stwo, we consider a vector $\vx \in \set{-1, 0, 1}^n$ such that 
$\vx_i=1$ if $i \in \Sone$, $\vx_i=-1$ if $i \in \Stwo$, and $\vx_i=0$ otherwise.

We now proceed to define our polarization measure:
given a community with two bands $(\Sone,\Stwo)$ we measure the degree of po\-lar\-iza\-tion in the community by
\begin{equation}  
  \label{eq:sbr}
  \begin{split}
    \sbr(\Sone, \Stwo) =\; 
    & \frac{2\abs{\posedges(\Sone, \Stwo)} + \abs{\negedges(\Sone)} + \abs{\negedges(\Stwo)}}{\vol(\Sone \cup \Stwo)} \\
    & +\frac{\abs{\edges(\Sone\cup\Stwo, \Srest)}}{\vol(\Sone \cup \Stwo)}.
\end{split}
\end{equation}
We refer to the quantity $\sbr$ as \textit{signed bi\-partite\-ness ratio}~\cite{atay2020cheeger}. 
% , which measures the degree of polarization 
% of the community $(\Sone, \Stwo)$.
In particular, the more polarized the community $(\Sone, \Stwo)$ 
the smaller the value of $\sbr(\Sone, \Stwo)$. 
Using $\vx$ (defined over $\Sone,\Stwo$) and $\vy=\Dhalf \vx$, the following relation holds:
\begin{equation}
  \label{eq:quad}
  \begin{split}
    \rquot(\vy) 
     =\; & \frac{\vx^T \lap \vx}{\vx^T \dmat \vx} \\
     =\; & \frac{ 4 \abs{\posedges(\Sone, \Stwo)} + 4 \abs{\negedges(\Sone)} + 4 \abs{\negedges(\Stwo)}}{\vol(\Sone \cup \Stwo)} \\
         & + \frac{ \abs{\edges(\Sone\cup\Stwo, \Srest)}}{\vol(\Sone \cup \Stwo)}.    
  \end{split}    
  % \rquot(\vy) = \frac{\vx^T \lap \vx}{\vx^T \dmat \vx} =  \frac{ 4 \abs{\posedges(\Sone, \Stwo)} + 4 \abs{\negedges(\Sone)} + 4 \abs{\negedges(\Stwo)} + \abs{\edges(\Sone\cup\Stwo, \Srest)}}{\vol(\Sone \cup \Stwo)}.
\end{equation}
Note that $\rquot(\vy)$ and $\sbr(\Sone, \Stwo)$ bound each other by a constant factor:
\begin{equation}
\label{eq:bound-each-other}
  \sbr(\Sone, \Stwo) \le \rquot(\vy) \le 4 \sbr(\Sone, \Stwo).
\end{equation}
Intuitively, minimizing $\sbr(\Sone, \Stwo)$ is closely related to minimizing the corresponding $\rquot(\vy)$.

As in the work of Atay and Liu~\cite{atay2020cheeger}, we refer to 
\[
  \cheegerc(\graph)=\min_{\Sone, \Stwo} \sbr(\Sone, \Stwo)
\]
as the \textit{signed Cheeger constant} of the graph \graph. 

\spara{Encoding information about seed nodes.}
%
% \cnote{Bruno: ``querying setting'' sounded a bit weird to me. Please let me know if you approve of this alternative.}
%
In this paper we consider a setting where a set of \emph{seed nodes} is given as input. 
We sometimes refer to seeds as \emph{query nodes}.
The seed nodes are specified as two disjoint sets $(\seedp, \seedn)$, 
i.e., $\seedp, \seedn \subseteq \nodes$ and $\seedp \cap \seedn = \emptyset$.
The sets \seedp and \seedn are meant to provide example nodes for the two bands of the polarized community 
we are interested in discovering.
Typically, $\abs{\seedp}$ and $\abs{\seedn}$ are small.
% Typically, the number of seeds is very small, e.g., $\abs{\seedp}$ and $\abs{\seedn}$ can be assumed to be small constants.
% \cnote{Bruno: "can be assumed to be small constants" Do we make use of this assumption anywhere?}

\cnote{Han: any interpretation on the denumerator $\vol(\seedp, \seedn)$ of the 3rd constraint below?}
\cnote{Aris: I added some explanation, not very long, as I think that it is self-explained.}
\cnote[Bruno]{Some intuition about the denominator: we want the solution to be close to the seeds, that is, preferably connected to them. The SBR favors density within the solution. Because of the ratio bound, every vertex in the solution that is not connected to the seeds is ``wasteful'', because it introduces $|S_1\cup S_2|$ units of sparsity. The denominator means that you can afford to add $k$ edges per edge in the seed, and neighbours are a usually better use of your budget.}

Intuitively, we are seeking a community $(\Sone, \Stwo)$ such that each band contains one of the seed sets, e.g., $\seedp \subseteq \Sone$ and $\seedn \subseteq \Stwo$ and at the same time minimizes $\sbr(\Sone, \Stwo)$.
In addition, we are interested in solutions whose size is bounded with respect to the size of the seed nodes and are preferably well connected to these.
We express this as a constraint of the volume of the set 
$\Sone\cup\Stwo$ normalized by the volume of the set $\seedp\cup\seedn$. 
In particular, we require that the volume of the solution $\vol(\Sone\cup\Stwo)$ is at most $k$ times
the volume of the seed nodes $\vol(\seedp\cup\seedn)$, where $k$ is a user-defined parameter.
We formalize this problem as follows:

\begin{problem*}[\localpoldiscreteshort]
  \label{prob:local-discrete}
  Given a signed graph $\graph$, a pair of seed sets $(\seedp, \seedn)$ and a positive number $k \in \real^{+}$
  our goal is to find a polarized community $(\Sone,\Stwo)$ so that
  \begin{enumerate}
  \item $(\Sone,\Stwo)$ minimizes the signed bi\-partite\-ness ratio $\sbr(\Sone,\Stwo)$;
  \item $\seedp \subseteq \Sone\text{ and } \seedn \subseteq \Stwo$;
  \item $\frac{\vol(\Sone \cup \Stwo)}{\vol(\seedp \cup \seedn)} \le k$;
  \end{enumerate}  
\end{problem*}
% \cnote{Han: can we say any hardness result of the above problem e.g., when $k=\infty$?}
% We include the denominator $\vol(\seedp \cup \seedn)$ in the second constraint for technical reasons which will be made clear later. 

Adapting the notion of signed Cheeger's constant $\cheegerc(\graph)$, 
we define a \textit{local variant} of $\cheegerc(\graph)$. 
Given a signed graph $\graph$, two disjoint node sets $\seedp$ and $\seedn$, and a positive number $k$, \textit{local signed Cheeger's constant} is defined as:

\begin{equation}
  \cheegerc(\seedp, \seedn, k) = 
  \!\!\!\!\!\!\!\!\min_{\substack{\Sone, \Stwo \subseteq \nodes: \\ \seedp \subseteq \Sone, \seedn \subseteq \Stwo \\ \vol({\Sone \cup \Stwo}) / \vol(\seedp\cup\seedn) \le k}} \!\!\!\!\!\!\!\!
  \sbr(\Sone, \Stwo)
\end{equation}

\smallskip
Next, we consider a continuous variant of the \localpoldiscreteshort problem. 
The theoretical properties of the continuous problem will be studied soon.
First, let $\vx$ be continuous, i.e., $\vx \in \real^{n}$.
Second, we encode the query information via a \textit{seed vector} $\vs \in \real^{n}$.
% a scheme used by Mahoney et al.~\cite{mahoney2012local}. 
Seed nodes take non-zero values in $\vs$ and signs indicate membership in \seedp, or \seedn:
by convention,  $\vs_i > 0$ if $i \in \seedp$, $\vs_i < 0$ if $i \in \seedn$, and $\vs_i = 0$ otherwise.
The magnitude of the coordinate $\vs_i$ indicates the strength by which we require that the solution contains
the node~$i$.
A special case of our setting is when one of the two query sets is empty, i.e., 
$\seedp = \emptyset$ or $\seedn = \emptyset$. 
For technical reasons, we normalize $\vs$ with respect to node degree, e.g., $\vs^T \dmat \vs = 1$.

We formalize the continuous problem below.
\begin{problem*}[\localpols]
  \label{prob:local}
  Given a signed graph $\graph$, a seed vector $\vs \in \real^n$,
  and a correlation parameter $\kappa$, our goal is to find a vector $\vx \in \real^n$ such that % is $\kappa$-correlated with \vs
  % and minimizes the signed bi\-partite\-ness ratio \sbr, that is
  % and a correlation parameter $\kappa$, our goal is to find a polarized community $(\Sone,\Stwo)$, 
  % encoded by a vector $\vx \in \set{-1, 0, 1}^n$, so that \vx is $\kappa$-correlated with \vs, 
  % and it optimizes the signed bi\-partite\-ness ratio \sbr:
  \begin{equation}
    \begin{aligned}
      & \underset{\vx}{\text{minimize}} && \vx^T \lap \vx \\
      & \textrm{such that} && \vx^T \dmat \vx = 1  \\
      &               && \vx^T \dmat \vs \ge \kappa  \\
      % &               && \vx \in \set{-1, 0, 1}^n.
      &               && \vx \in \real^n.
    \end{aligned}
  \end{equation}
\end{problem*}

The parameter $\kappa$ controls the correlation between $\vx$ and $\vs$.
Notice that \localpols is not convex 
due to the constraint $\vx^T \dmat \vx = 1$. % is not convex.

It can be shown that \localpols is indeed a \emph{continuous relaxation} of \localpoldiscreteshort.
This is shown in Claim~\ref{claim:relaxation}, below.
Using this result, we can solve an instance of the \localpoldiscreteshort problem, 
by first solving its continuous relaxation and then rounding the solution to a discrete one.
The result is also used to prove the approximation guarantee for \localpoldiscreteshort.

% \cnote{Han: added the purpose of Claim 1}
% \cnote{Aris: OK, good, I edited a bit.}

Before stating the claim, we introduce some additional notation.
For $S \subseteq \nodes$,
we let $1_S \in \set{0, 1}^n$ be a vector that has 1 at entries in $S$
and 0 otherwise.
Then, for two disjoint sets $A, B \subseteq \nodes$,
we define the vector $\vsAB = (\vA - \vB) / \sqrt{\vol(A \cup B)}$. 
It can be checked that $\vsAB^T\;\dmat\;\vsAB = 1$.

\begin{claim}
  \label{claim:relaxation}
  For any $\seedp, \seedn \subseteq \nodes$, \localpolk is a relaxation of \localpoldiscrete. 
  In particular,
  \begin{equation}
    \label{ineq:relax}
    \lambda(\vsuv, \sqrt{1/k}) \le 4 \cheegerc(\seedp, \seedn, k),
  \end{equation}
  where
  $\lambda(\vsuv, \sqrt{1/k})$ is the optimal value of \localpolk
  and $\cheegerc(\seedp, \seedn, k)$ is the optimal value of \localpoldiscrete. 
\end{claim}
\begin{proof}
  First, the objective of \localpols relaxes the objective of \localpoldiscreteshort
  because \localpols optimizes over a continuous space and the two objectives bound each other up to constant factors (Equation~\ref{eq:bound-each-other}).
  
  Second, \localpols relaxes both constraints of \localpoldiscreteshort and translates them into the correlation constraint.
  Consider any feasible solution  $\Sone$ and $\Stwo$ of \localpoldiscreteshort, and let $C=\Sone \cup \Stwo$ and $S=\seedp\cup\seedn$.
  Also let $\vx = \vxab$ and $\vs = \vsuv$ to be used in \localpols.
  Expanding $\vx^T \dmat \vs$ gives:
  \newcommand{\denum}{\ensuremath{\frac{1}{\sqrt{\vol(C)\vol(S)}}}\xspace}
  \begin{align*}
    \vx^T \dmat \vs &= \denum (\vone - \vtwo)^T \dmat (\vu - \vv) \\
                &= \denum (\vone^T \dmat \vu  + \vtwo^T \dmat \vv) \\
                &= \denum (\vol(\seedp) + \vol(\seedn)) \\
                &= \sqrt{\frac{\vol(S)}{\vol(C)}} \ge \sqrt{\frac{1}{k}}.
  \end{align*}
  Line 2 follows from: $\Sone \cap \Stwo = \emptyset$, $\seedp \cap \seedn = \emptyset$ and $\seedp\subseteq\Sone, \seedn\subseteq\Stwo$.

  Last, it is easy to verify $\vx^T \lap \vx \le  4 \beta(\Sone, \Stwo)$.
  % \begin{align*}
  %   \vx^T \lap \vx = & \frac{ 4 \abs{\posedges(\Sone, \Stwo)} + 4 \abs{\negedges(\Sone)} + 4 \abs{\negedges(\Stwo)}}{\vol({C})} \\
  %   & + \frac{\abs{\edges(\Sone \cup \Stwo, \Srest)}}{\vol({C})} \\
  %   \le & 4 \beta(\Sone, \Stwo) 
  %     \end{align*}  
Also by definition, we have $\lambda(\vsuv, \sqrt{1/k}) \le \vx^T \lap \vx$.
So it follows immediately that $\lambda(\vsuv, \sqrt{1/k}) \le 4 \cheegerc(\seedp, \seedn, k)$.
\end{proof}

% We rely on solving \localpols in order to approximate \localpoldiscreteshort. 
In the next section, we analyze the properties of \localpols and propose an approximation algorithm for \localpoldiscreteshort.

% \cnote{Aris: it was $\vx \in \real^n$ and I changed it to $\vx \in \set{-1, 0, 1}^n$. We should provide discussion about this. }
% \cnote{Bruno: I think we're mixing two things here, because we cannot enforce $\vx^T \dmat \vx = 1$ for $\vx \in \set{-1, 0, 1}^n$, right? The discrete problem should not have that constraint, and the correlation constraint can be hard, e.g., seed $\subseteq C_1\cup C_2$ (wasn't it like that before?)}

% \cnote{Aris: it fact it is a discrete problem. Should we also define the continuous relaxation? }

\section{Theoretical analysis}

% \cnote{add the high-level proof viz and simply it a bit }
% \cnote{Aris: looks ok to me.}

We first present how to solve the continuous problem \localpols and obtain an optimal solution $\xopt$. 
This is stated in Theorem~\ref{thm:solution}.
Then, we describe how to round the continuous optimal solution $\xopt$ so as to obtain a discrete solution to 
\localpoldiscreteshort with approximation ratio $\bigO(\sqrt{\cheegerc(\seedp, \seedn, k)})$.
Finally, we present an efficient algorithm \ours (Algorithm~\ref{alg}) that leverages these~results. 

% Next we present main results related to \localpols.

% \paragraph{Solution characterization.}
% We first characterize the solution of \localpols in closed-form.

% \cnote{Todo: add more proof details if necessary}

\begin{theorem}[Solution characterization]
  \label{thm:solution}
%  \textbf{.}
  Given an unbalanced signed graph $\graph$ with normalized Laplacian matrix $\nlap$, 
  let $\vs \in \real^{n}$ be a seed vector, with $\vs^T \dmat \vs = 1$ and $\vs^T \dmat \evec \neq 0$, where $\evec$ is the eigenvector of the \textit{smallest} eigenvalue of $\nlap$.
  % the first eigenvector of $\nlap$.
  Let $0 \le \kappa < 1$ be a correlation parameter, and let $\xopt$ be an optimal solution to $\localpol$. Then, there exists some $\alpha \in (-\infty, \eigval)$ and a $\beta \in [0, \infty]$ such that
  \[
    \xopt = \beta \pinv{\lap - \alpha \dmat} \dmat \vs
  \]
  where $\pinv{\lap - \alpha \dmat}$ is the pseudo inverse of $\lap - \alpha \dmat$.
\end{theorem}

% We describe a  proof sketch of Theorem~\ref{thm:solution} in the next section and leave the remaining details in appendix. 

\begin{theorem}[Approximation guarantee]
  \label{thm:appx}
%  \textbf{Approximation guarantee.}
  Given an unbalanced signed graph $\graph$, two disjoint node sets $\seedp, \seedn \in \nodes$, and a positive integer $k$, we can find two disjoint sets $\Sone$ and $\Stwo$ that achieve $\sbr(\Sone, \Stwo) = \bigO(\sqrt{\cheegerc(\seedp, \seedn, k)})$ from the optimal solution of \localpolk.
  % Moreover, the whole procedure can be done in nearly-linear time in the size of the graph.
\end{theorem}

% We leave the proof of Theorem~\ref{thm:appx} in the appendix as it is relatively trivial.

% We will describe how to efficiently find $\alpha$ later.

% \paragraph{Approximation guarantee.} 
% We show it is possible to find two sets with a good approximation ratio w.r.t $\sbr$.

% We first define some notation. 
% For $S \subseteq \nodes$,
% we let $1_S \in \set{0, 1}^n$ be a vector which is 1 for vertices in $S$
% and 0 otherwise.
% Then for two disjoint sets $\Sone, \Stwo \subseteq \nodes$ and $C=\Sone \cup \Stwo$,
% we define the vector $\vxab = (\vone - \vtwo) / \sqrt{\vol(C)}$. 
% It can be checked that $\vxab^T \dmat \vxab = 1$. 
% If $\Sone=\set{u}, \Stwo=\set{v}$,
% we denote $\vsuv$ as a shorthand of $\mathbf{x}_{\set{u}, \set{v}}$. 

\begin{figure*}[t]
  \resizebox{0.9\linewidth}{!}{
    \begin{tikzpicture}
      % \draw[help lines] (1,1) grid (15,4);
      \tikzstyle{line} = [line width=0.05cm, draw=gray!60]
      \tikzstyle{box}  = [inner sep=2pt, draw=white, outer sep=0]

      \node[draw, box, rectangle] at (2,4) (discrete) {
        \localpoldiscreteshort
      };
      
      \node[draw, box, rectangle] at (8,4) (spectral) {
        \localpols
      };
      \node[draw, box, rectangle] at (13,4) (sdp) {
        \sdpps
      };

      \node[draw, box, rectangle, text width=4cm] at (13,3) (optimality) {
        \textbf{Solution characterization}  (Theorem~\ref{thm:solution}) 
        % 1. $\rank(\Xopt)=1$  (Claim~\ref{claim:rank1}) \\
        % 2. closed-form $\xopt$ (Claim~\ref{claim:closed-form})
      };  

      \node[draw, box,  rectangle, text width=2.5cm] at (3,3) (lb) {
        \textbf{Relaxation result \\ }
        (Claim~\ref{claim:relaxation})\\
      };
      
      % Cheeger's part
      \node[draw, box,  rectangle, text width=4.2cm] at (8,3) (sweeping) {
        \textbf{Signed Cheeger's inequality \\}
        (Proposition~\ref{pp:signed-sweep}, Atay et al.~\cite{atay2020cheeger})\\
      };

      \node[draw, box,  rectangle, text width=6cm] at (8,2) (final) {
        \textbf{Approximation guarantee} 
        (Theorem~\ref{thm:appx}) 
      };
      
      % \draw [->] (problem) -- node[right] {transform + relax} (spectral);
      \draw [->, line] (discrete) -- node[above] (first_relax) {relaxes to} (spectral);
      \draw [->, line] (spectral) -- node[above] {relaxes to} (sdp);
      \draw [->, line] (spectral) -- node[below] {(zero optimality gap)} (sdp);
      \draw [->, line] (first_relax) -- (lb);
      % \draw [->, line] (spectral) -- (lb);
      % \draw [->, line] (sdp) -- node[] {duality \quad theory} (optimality);
      \draw [->, line] (sdp) -- node[] {} (optimality);
      % \draw [->] (sdp) -- node[below] { theory} (optimality);
      \draw [->, line] (optimality) -- (final);
      \draw [->, line] (sweeping) -- (final);
      \draw [->, line] (lb) -- (final);
    \end{tikzpicture}
  }
  \caption{High-level structure of our proof.}
  \label{fig:high-level-proof}
\end{figure*}
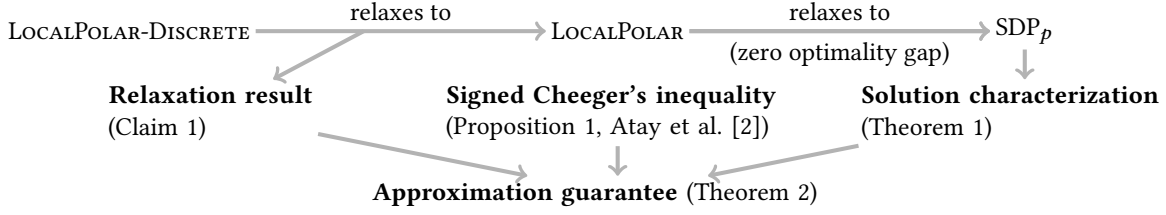

% \cnote{Han: the above notations are not used in the theorem, move them to proofs}
We illustrate the high-level structure of our proof in Figure~\ref{fig:high-level-proof}.

\label{sec:proof}
\subsection{Proof of Theorem~\ref{thm:solution}}

% Our proof for Theorem~\ref{thm:solution} is carried out as follows:
We first relax \localpol to a convex semi-definite program \sdpp shown in Figure~\ref{fig:sdp-relax}.
Then we show that the optimality gap between \sdpps and \localpols is zero.\footnote{ We omit problem arguments when the context is clear, e.g., we use \sdpps for brevity.}
By using the complementary slackness conditions of \sdpps and its dual % (in Figure~\ref{fig:sdp-relax}),
we show that the solution of \sdpps has rank 1, which characterizes its form.
Our proof technique is adapted from Mahoney et al.~\cite{mahoney2012local}.

\begin{figure}[h!]
  \begin{center}
    \begin{tabular}{cc}
      Primal \sdpp  & Dual \sdpd \\
      {$\begin{aligned}
          & \min_{X} && \lap \mprod X  \\
          & \textrm{s.t.} && \dmat \mprod X = 1  \\
          &               && (\dmat \vs)(\dmat \vs)^T \mprod X \ge \kappa  \\
          &               && X \geneq 0  
        \end{aligned}$} &
                          {$\begin{aligned}
                              & \max_{\alpha, \beta} && \alpha + \beta \kappa  \\
                              & \textrm{s.t.} &&  \lap \geneq \alpha \dmat + \beta (\dmat\vs)(\dmat\vs)^T \\
                              &               &&  \beta \ge 0 \\
                              &               && \alpha \in \real
                            \end{aligned}$}
    \end{tabular}
  \end{center}
  \caption{SDP relaxation of \localpol (left) and its dual problem (right)}
  \label{fig:sdp-relax}
\end{figure}

To see that \sdpps is a relaxation of \localpols, consider a feasible solution $x$ of \localpols. Then $X=xx^T$ is also a feasible solution for \sdpps and $X \mprod \lap=\vx^T \lap \vx$.

The following two claims are based on known results in convex optimization~\cite{boyd2004convex}:

\begin{claim}
  Strong duality holds between \sdpp and \sdpd. 
\end{claim}
\begin{proof}
  As \sdpp is a convex problem, Slater's condition is sufficient to verify zero duality gap~\cite{boyd2004convex} (Chapter 5.5.3).
  By setting $X=\vs \vs^T$, we have $(\dmat \vs)(\dmat \vs)^T \mprod \vs \vs^T = (\vs^T \dmat \vs)^2 = 1 > \kappa$, i.e., strong feasibility holds. Therefore Slater's condition is satisfied. 
\end{proof}

\begin{claim}
  \label{claim:strong-duality-and-conditions}
  Strong duality holds between \sdpp and \sdpd. 
  Therefore, the feasibility and complementary slackness conditions suffice to establish optimality.

  The feasibility conditions are:
  \begin{align}
    \label{eq:feas1} \dmat  \mprod \Xopt & = 1,  \\
    \dsds \mprod \Xopt & \ge \kappa, \nonumber \\
    \Xopt &\geneq 0 \nonumber \\
    \label{eq:feas2}\lap - \alphaopt \dmat - \betaopt \dsds & \geneq  0, \: and \\
    \betaopt & \ge 0 \nonumber
  \end{align}
  and complementary slackness conditions are:
  \begin{align}
    \label{eq:cs1} \betaopt (\dsds \mprod \Xopt  - \kappa) & = 0, \: and \\
    \label{eq:cs2} \Xopt \mprod (\lap - \alphaopt \dmat - \betaopt \dsds) &= 0
  \end{align}  
\end{claim}
\begin{proof}
  These conditions directly follow from the optimality conditions of semi-definite programs~\cite{boyd2004convex} (Chapter 5.9.2). 
\end{proof}

A key finding is the following claim. The proof is given in the Appendix.

\begin{claim}
  \label{claim:rank1}
  The rank of $\Xopt$ is 1, i.e, $\Xopt=\xopt (\xopt)^T$, and therefore the optimality gap between \localpol and \sdpp is~0. 
\end{claim}

\begin{claim}
  \label{claim:closed-form}
  Given that $\Xopt$ has the form $\xopt (\xopt)^T$, the feasibility and complementary slackness conditions imply,
  \[
    \xopt = \pm \betaopt \sqrt{\kappa} \pinv{\lap - \alphaopt \dmat} \dmat\vs .
  \]
\end{claim}
\begin{proof}
  Using Claim~\ref{claim:rank1}, we have $\Xopt=\xopt (\xopt)^T$. Relying on equation~(\ref{eq:feas2})
  we can rewrite Equation~(\ref{eq:cs2}) as:
  \begin{align*}
    (\lap - \alphaopt \dmat - \betaopt \dsds) \xopt &= 0 \\
    % (\lap - \alphaopt \dmat) \xopt - \betaopt \dsds \xopt &= 0 \\
    (\lap - \alphaopt \dmat) \xopt  &= \pm \betaopt  \sqrt{\kappa} \dmat \vs \\
    \xopt &= \pm \betaopt \sqrt{\kappa} \pinv{\lap - \alphaopt \dmat} \dmat\vs .
  \end{align*}
  The 2nd line follows from Equation~(\ref{eq:cs1}), as $(\dmat \vs)^T \xopt = \pm \sqrt{\kappa}$. 
\end{proof}

% Based the conditions in Claim~\ref{claim:strong-duality-and-conditions}, we have the following:

% \begin{claim}
%   \label{claim:rank1-and-closed-form}
%   $\rank(\Xopt)=1$, therefore the optimality gap between \localpols and \sdpps is zero.
%   Combining $\rank(\Xopt)=1$ with feasibility and complementary slackness conditions implies,
%   \[
%     \xopt = \pm \betaopt \sqrt{\kappa} \pinv{\lap - \alphaopt \dmat} \dmat\vs
%   \]
% \end{claim}

% Though \localpol can now be solved by \sdpp, which is a convex problem, 
% we next give a closed-form of $\Xopt$ which is easier to compute. 

% In supplementary material, we describe a procedure to approximate $\vxopt$ in $\bigO(\sqrt{c}m\log(m))$, where $c$ is the condition number of $\lap - \alphaopt \dmat $. 

\subsection{Proof of Theorem~\ref{thm:appx}}
Our proof relies on Cheeger's inequality for signed graphs:

% \cnote{Pin-pointed which lemma in the original paper and say that the result below is a special case of the original work?}

\begin{proposition}
  \label{pp:signed-sweep}
  (Lemma 4.2 by Atay et al.~\cite{atay2020cheeger})
  \footnote{Our statement is a special case of the original lemma: we set $\mu$ to be node degree.
    Also, we use $t$ instead of $\sqrt{t}$. It's easy to see they're equivalent. }
  For any non-zero vector $\vx \in \real^{\abs{V}}$, there exists a $t \in [0, \max_{u \in \nodes} \abs{\vx_u}]$ such that
  \[    
    \sbr(\nodes_{\vx}(t), \nodes_{\vx}(-t)) \le \sqrt{2 \frac{\vx^T \lap \vx}{\vx^T \dmat \vx}}
  \]
  where $\nodes_{\vx}(t) = \set{u \in \nodes \mid \vx_u \ge t} \text{ and } \nodes_{\vx}(-t) = \set{u \in \nodes \mid \vx_u \le -t}$.
\end{proposition}

\paragraph{Proof of Theorem~\ref{thm:appx}.}
Given a problem instance of \localpols with parameters $\vsuv$ and $\kappa=\sqrt{1/k}$, we apply Theorem~\ref{thm:solution} and get the optimal vector $\xopt$.
By Proposition~\ref{pp:signed-sweep}, we can get two sets $\Sone$ and $\Stwo$ that have $\sbr(\Sone, \Stwo) \le \sqrt{2\lambda(\vsuv, \sqrt{1/k})}$. 
Finally, by Claim~\ref{claim:relaxation}, $\sbr(\Sone, \Stwo) = \bigO(\sqrt{\cheegerc(\Sone, \Stwo, k)})$. 

\subsection{\ours algorithm}

We now summarize the complete algorithm, which we call \ours. 
Pseudocode is presented as Algorithm~\ref{alg}.
The \ours algorithm consists of 2 steps:
($i$) approximate within small error the optimal solution $\xopt$ of the problem \sdpps, 
using Theorem~\ref{thm:solution}; and
($ii$)~round $\xopt$ to obtain an integral solution, using Proposition~\ref{pp:signed-sweep}.
We discuss these two steps in more detail.

\begin{small}
\begin{algorithm}[t]
  \SetAlgoLined
  \KwData{
    Signed graph $\graph=(\nodes, \posedges, \negedges)$,
    seeds $(\seedp, \seedn)$,
    $k \in \real^{+}$ and 
    tolerance parameter $\tol>0$
  }
  \KwResult{two bands $(\Sone, \Stwo)$}
  Create seed vector $\vs$ s.t. $\vs_i=1 \text{ if } i\in\Sone$ and $\vs_i=-1 \text{ if } i\in\Stwo$\;
  normalize $\vs$ s.t. $\vs^T \dmat \vs=1$\;\label{alg:part1_start}
  $\kappa=1/\sqrt{k}, \ \ell  \define -\vol(\nodes), \ u \define \eigval$\;\label{alg:bs1}
  \Repeat{$-\tol \le \sqrt{\kappa} - \vs^T \dmat \vx \le 0$}{
    $\alpha \define (\ell +u)/2$\;
    Solve $(\lap - \alpha \dmat)\vx= \dmat\vs$ using Conjugate Gradient method\;
    normalize $\vx$ s.t. $\vx^T \dmat \vx=1$\; 
    \lIf{$\sqrt{\kappa} - \vs^T \dmat \vx<0$}{$\ell  \define \alpha$}\lElse{$u \define \alpha$}
  }\label{alg:bs2}
  % Find $t \in \real^+$ that minimizes $\sbr(\nodes_{\vx}(t), \nodes_{\vx}(-t))$, where $\nodes_{\vx}(t) = \set{u \in \nodes \mid \vx_u \ge t}$ and $\nodes_{\vx}(-t) = \set{u \in \nodes \mid \vx_u \le -t}$\;\label{alg:rounding1}
  Find $t \in \real^+$ that minimizes $\sbr(\nodes_{\vx}(t), \nodes_{\vx}(-t))$\;\label{alg:rounding1}  
  \Return{$(\nodes_{\vx}(t), \nodes_{\vx}(-t))$}
  \caption{
    \ours.
    The algorithm first approximates $\xopt$ by performing binary search on $\alpha$ (lines~\ref{alg:bs1}--\ref{alg:bs2}).
    Then, $(\Sone, \Stwo)$ is attained by rounding \vx at line~\ref{alg:rounding1}. An $\bigO(m+n\log n)$ rounding procedure is described in Appendix~\ref{appendix:fast-sweep}.
    % In experiment, we set $\epsilon=10^{-3}$. 
  }
  \label{alg}
\end{algorithm}
\end{small}

\spara{Computing the optimal solution $\xopt$.}
We discuss how to find the optimal solution $\xopt = \betaopt \pinv{\lap - \alphaopt \dmat} \dmat \vs$ without knowing the value of $\alphaopt$ beforehand.\footnote{Note that once $\alphaopt$ is found, the value $\betaopt$ can be computed by normalizing $\pinv{\lap - \alphaopt \dmat} \dmat \vs$ with respect to $\dmat$.}
If we assume that $\alphaopt$ is known, then $\pinv{\lap - \alphaopt \dmat}$ 
can be approximated using Conjugate Gradient Method~\cite{golub2012matrix}
in time $\bigO({m \sqrt{c}})$, where $c$ is the condition number of $\lap - \alphaopt \dmat$.

Then, to approximate $\alphaopt$, we use complementary slackness condition Equation~(\ref{eq:cs1})
% $(\vxopt)^T (\dmat \vs)(\dmat \vs)^T \vxopt = \kappa$ % 
and apply binary search on $\alpha \in [-\vol(\graph), \eigval)$. 
The search stops when $\vx^T (\dmat \vs)(\dmat \vs)^T \vx$ is close enough to $\kappa$ (controlled by $\tol$).
The whole process requires time $\bigO(\sqrt{c}m\log(m))$. 

Lines~\ref{alg:bs1} to~\ref{alg:bs2} in Algorithm~\ref{alg} describe the above process. 

\spara{Rounding $\xopt$.}
% \label{sec:efficient-rounding-brief}
We want to find a value  $t \in \real^{+}$ that minimizes $\sbr(\nodes_{\vx}(t), \nodes_{\vx}(-t))$. % (described in Proposition~\ref{pp:signed-sweep})
A na\"ive way is to try each value in $\vx$ for $t$, 
compute the corresponding $\sbr(\nodes_{\vx}(t), \nodes_{\vx}(-t))$ \textit{from scratch} 
(doable in time $\bigO(m)$) and take the value of $t$ that minimizes \sbr.
The total runtime cost is $\bigO(mn+n\log(n))$.

We propose a faster procedure, which \textit{incrementally} computes
$\sbr(\nodes_{\vx}(t), \nodes_{\vx}(-t))$ for values of $t$ in ascending order.
The time cost reduces to $\bigO(m+n\log(n))$. 
Moreover, our implementation leverages basic vector and matrix operations, which further boost speed in practice.
The details are described in Appendix~\ref{appendix:fast-sweep}.
\section{Experimental evaluation}

In this section we evaluate the performance of the proposed algorithm.
Our experimental evaluation is structured as follows.
First in Section~\ref{subsec:synthetic},
we experiment with synthetic graphs to better understand the behavior of our method under different scenarios.  
In Section~\ref{subsec:comparative}, we compare with a state-of-the-art method~\cite{chu2016finding} for polarized community detection on real-world graphs. 
In Section~\ref{subsec:casestudy} we discuss two case studies, and finally, 
in Section~\ref{subsec:scalability} we demonstrate the scalability of our algorithm.

We implement our algorithm in Python and perform the experiment on a Linux machine 
with a 10-core CPU and 64\,GB of memory.
We use the Conjugate Gradient solver implemented in \href{https://docs.scipy.org/doc/scipy-1.2.1/reference/}{Scipy 1.2.1} with the default parameters.
We set the binary search parameter $\tol=10^{-3}$ (used by Algorithm~\ref{alg}).
\cnote{Cite Aalto Triton}

The implementation of the code will be available in the non-anonymized version of the paper.

% \cnote{Todo: more technical details on linear system solver, e.g., number of iterations, tolerance parameter, etc }

\begin{table}[h]
  \centering
  \caption{Real-word signed graphs: $\eigval$ is the smallest eigenvalue of the normalized Laplacian. }
  \label{tbl:dataset}
  \resizebox{\columnwidth}{!}{
    \begin{tabular}{lrrrrr}
      \toprule
      Signed graph &   $\abs{\nodes}$ &  $\abs{\edges}$ &  $\abs{\negedges} / \abs{\edges}$ &  $\eigval$ & Category \\
      \midrule
      \dsword &    5\,K &      47\,K &  0.20 &  0.030 & language  \\
      \dsbc   &    6\,K &      21\,K &  0.15 &  0.040 & social  \\
      \dsref  &   11\,K &     251\,K &  0.05 &  0.039 & political\\
      \dssd   &   82\,K &     500\,K &  0.24 &  0.017 &  social \\
      \dsep   &  119\,K &     704\,K &  0.17 &  0.011 &  social \\
      \dswiki &  113\,K &  2\,000\,K &  0.63 &  0.075 &  edit conflict  \\
      \midrule 
      \dswikilarge &  1\,M &  33\,M &  0.63  &  0.075 & artificial \\ 
      \bottomrule
    \end{tabular}
  }
\end{table}

\spara{Real-world datasets.}
We choose publicly-available real-world signed graphs, whose statistics are summarized in Table~\ref{tbl:dataset}.
\dsword~\cite{sedoc2017semantic} captures the synonym-and-antonym relation among words in the English language.
A synonym (resp.\ antonym) pair is connected by an edge with positive (resp.\ negative) weight, 
where a weight represents the strength of similarity (resp.\ opposition).
\dsref~\cite{lai2018stance} records the 2016 Italian Referendum on Twitter:
an edge is negative if two users are classified with different stances, and positive otherwise.
\dsbc is a trust network of Bitcoin users. 
\dssd\footnote{http://snap.stanford.edu/data/soc-Slashdot0902.html}
is a social network from a technology-news website,
which contains friend or foe links among its users.
\dsep\footnote{http://snap.stanford.edu/data/soc-Epinions1.html} is a trust online social network from a consumer review site, where site members can decide to trust, or distrust, each other.
Edges in \dswiki\footnote{http://konect.uni-koblenz.de/networks/wikiconflict} represents positive and negative edit conflicts between users on Wikipedia.

To evaluate the scalability of our method, 
we artificially augment \dswiki to \dswikilarge.  % so that it contains $1$M nodes
% . We denote the augmented version as \dswikilarge.
The procedure is described in Section~\ref{subsec:scalability}. 
We experiment with the largest weakly-connected component of each graph. 
Directed graphs are converted to undirected ones with adjacency matrix $(\adjm + \adjm^{T}) / 2$.

\begin{figure*}[h]
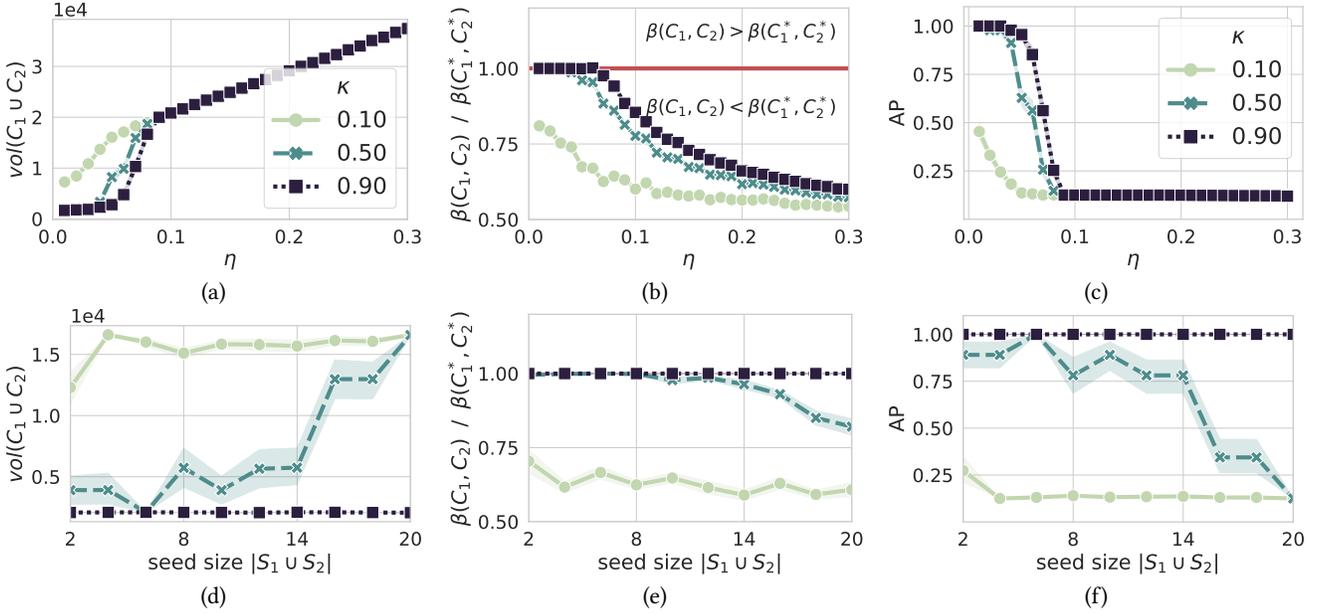

  \centering
  \newcommand{\addfigure}[1]{\includegraphics[width=0.31\linewidth]{synthetic_experiment/#1.pdf}}  
  \begin{tabular}{ccc}
    \addfigure{{effect_of_eta_solution_size}} &  \addfigure{{effect_of_eta_beta_ratio}} & \addfigure{{effect_of_eta_ap}} \\
    (a) & (b) & (c)\\
    \addfigure{{effect_of_seed_solution_size}}  &  \addfigure{{effect_of_seed_beta_ratio}} & \addfigure{{effect_of_seed_ap}} \\
    (d)  & (e) & (f) \\
  \end{tabular}
  \caption{
    % \inote{Fix the caption because new row of sub figures}
    Performance of different metrics on synthetic graphs.
    1st/2nd/3rd columns corresponds to : solution volume / \sbr ratio / average precision respectively.
    % \textbf{2nd column}: \sbr ratio, $\sbr(\Sone, \Stwo) / \sbr(\Sonetrue, \Stwotrue)$.
    % and \textbf{3rd column}: average precision (\ap);
    % and \textbf{bottom}: solution size, $\abs{\Sone \cup \Stwo}$.
    Meanwhile, we vary 2 parameters:
    (i) edge noise parameter $\noise$ (top row);
    (ii) seed size $\seedcount$ for $\Sonetrue$ and $\Stwotrue$ (bottom row).
    Results are shown for 3 values of the parameter $\kappa$.
    % For each configuration, the result is obtained by averaging over 10 random graphs and 32 random seeds.
    In figure (b), we use a red solid line at $y=1$  to differentiate two cases of $\sbr$ ratio.
    Error bars at 0.95 confidence interval is drawn in the form of shaded area. 
    % In figure (g)-(i), solid red lines are drawn horizontally at $\abs{\Sonetrue \cup \Stwotrue}$.
  }
  \label{fig:syn}
\end{figure*}

% \cnote{Han: I removed the parameter community number and community size since they're fixed.}

\spara{Synthetic datasets.}
To better assess the effectiveness of
our method under different scenarios, 
we create random graphs containing 8 ground-truth polarized communities, 
$\left\{(C_{i,1},C_{i,2}) \mid i=1,\dots, 8\right\}$. % , surrounded by outlier nodes.
For simplicity, we assume that all bands are of the same size, $20$.
% The creation process is controlled by the following additional parameters:
% $\commsize = \abs{\Sione} = \abs{\Sitwo}$, for $i=1, \ldots, \commcount$, is the size of each % polarized
% band 
% ;
In addition, the creation process is controlled by an edge-noise parameter $\noise \in [0, 1]$,
where large values of $\noise$ indicate more noise in edge signs. 
% In our experiments, we fix $\commsize=20$ and $\commcount=8$, and vary the other parameters.
The role of edge-noise parameter $\noise$ is as follows:

\squishlist
\item 
for each pair of nodes within \Sione or \Sitwo, for $i=1, \ldots, \commcount$,
there is a positive edge with probability $1-\noise$, 
a negative edge with probability $\noise / 2$, 
and non edge with probability $\noise / 2$;
\item 
for each pair of nodes, one in \Sione and the other in \Sitwo, for $i=1, \ldots, \commcount$,
there is a negative edge with probability $1-\noise$,
a positive edge with probability $\noise / 2$, 
and a non edge with probability $\noise / 2$;
\item for every other pair of nodes in the graph
there is a non edge with probability $1-\noise$,
a positive edge with probability $\noise / 2$,
and a negative edge with probability $\noise / 2$.
\end{list}

In our experiments, we also vary the size of seed set $\abs{\seedp \cup \seedn}$ and the value of $\kappa$.
For simplicity, we make $\abs{\seedp} = \abs{\seedn}$. 

\spara{Baseline.}
We compare our method with a state-of-the-art method on discovering polarized communities on real-world graphs.
The \focg\ algorithm~\cite{chu2016finding} is designed to discover $k$-way polarized communities. 
% which are communities having mostly positive edges within bands and mostly negative edges across bands. 
Note that \focg\ is a global method, and thus, not directly comparable with our approach; 
we select \focg\ as baseline as it is the most related work in the literature. % with our problem formulation. 
% \cnote{It seems to me that we are using groups to refer to either $k$-way polarized structures and each of the $k$ subgraphs interchangeably. This is confusing, so let's make sure we distinguish both with different names.}
% Given a signed graph,
\focg returns a set of $k$-way polarized communities,
$\set{\mathcal{C}_i}_{i=1}^\ell$, where $\mathcal{C}_i=(C_{i1}, \ldots, C_{ik})$ and $\ell$ is determined by the algorithm.
\focg ensures that $\mathcal{C}_i \cap \mathcal{C}_j = \emptyset$ for all $i\neq j$.
% In our experiments
We use $k=2$ and set the other parameters as suggested.
% To align better with our setting
We exclude communities that either
($i$) are too small, in particular, $\abs{\Sone} \text{ or } \abs{\Stwo}\le 5$, or 
($ii$) the two bands intersect.% , i.e., $\Sone \cap \Stwo \neq \emptyset$. 

\subsection{Evaluation on synthetic graphs}
\label{subsec:synthetic}

% \cnote{Han: I removed the results on ``effect of outlier nodes'' as I find things are becoming complicated in this situation --  I'm not able to interpret the results to a satisfactory degree.}

% To evaluate \ours more comprehensively,
By studying the effects of various parameters including algorithm and graph generation parameters, we aim to answer the following research questions:

\squishdesc
% \begin{itemize}
\item \textbf{Q1:} what is the effect of parameter $\kappa$ on the characteristics of the solution, as well as in the accuracy of finding the ground truth?
\item \textbf{Q2:} what is the effect of noise in the behavior of \ours?
\item \textbf{Q3:} can we find a polarized community more accurately by giving more seeds to \ours?
% \end{enumerate}
\end{list}

We report the average performance for each configuration over a combination of 10 randomly generated graphs and 10 randomly selected ground-truth communities paired with random seeds in them.
% Performance for each parameter configuration is averaged over a combination of 10 randomly generated graphs and 32 seeding rounds.
% We describe the randomization method in supplementary material. 

\spara{Evaluation metric.}
% We treat the problem of finding the true pair $(\Sonetrue, \Stwotrue)$ as a multi-class classification problem.
% , where the true node labeling is
% $\nodelabel(v)=+1$ if $v \in \Sonetrue$, $\nodelabel(u)=-1$ if $u \in \Stwotrue$, and $\nodelabel(u)=0$ otherwise.
Given a solution $(\Sone, \Stwo)$ and ground truth $(\Sonetrue, \Stwotrue)$, we measure its \emph{average precision}.%
\footnote{W.l.o.g., we omit the issue of label permutation in the definition of~$\ap$.}
% \[\ap(\Sone, \Stwo) = \left( \abs{\Sone \cap \Sonetrue} / \abs{\Sone} + \abs{\Stwo \cap \Stwotrue} / \abs{\Stwo} \right) / 2.\]
\[
  \ap(\Sone, \Stwo) = 
  	\frac{1}{2} \left( \frac{\abs{\Sone \cap \Sonetrue}}{\abs{\Sone}} +  
  	\frac{\abs{\Stwo \cap \Stwotrue}}{\abs{\Stwo}} \right).
\]

% To better understand the performance of \ap, 
We also report the ratio $\sbr(\Sone, \Stwo) / \sbr(\Sonetrue, \Stwotrue)$, referred as  \emph{$\sbr$-ratio} for brevity.
The motivation is to see whether our method can find a solution that is even better than the ground truth, according to the optimized measure, which happens when $\sbr$ ratio $<1$.
% the solution found by our method is better than ground truth.
Finally, we report the solution volume $\vol(\Sone \cup \Stwo)$. 

\begin{figure*}[h]
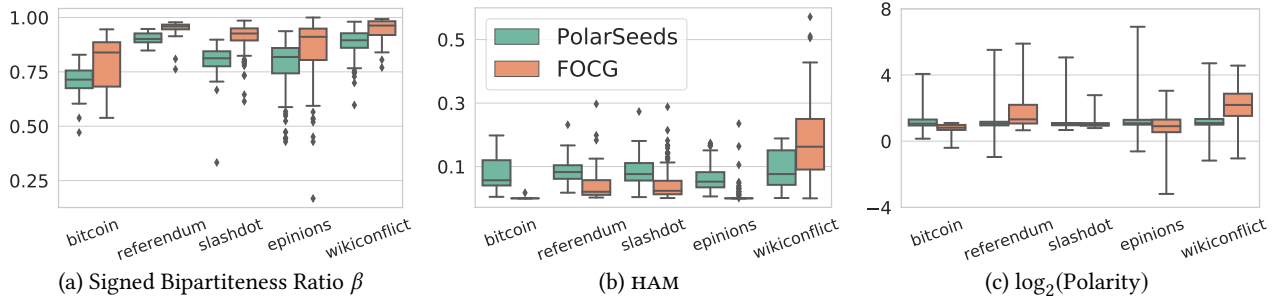

  \centering
  \newcommand{\addfigure}[1]{\includegraphics[width=0.30\linewidth]{real_graph_exp/#1.pdf}}  
  \begin{tabular}{ccc}  
    \addfigure{{beta_distribution}} & \addfigure{{ham_distribution}} & \addfigure{{pc_distribution}} \\
    (a) Signed Bipartiteness Ratio \sbr & (b) \ham & (c) $\log_2(\text{\pc})$
  \end{tabular}
  \caption{
    Distributions of 3 evaluation metrics on all communities found by \ours and \focg.
    \focg returns multiple communities by design.    
    We randomly seed \ours multiple rounds  in order to find multiple communities.
    Overlapping communities are filtered out to make the comparison fair.
    % We designed the process so that \ours and \focg find multiple non-overlapping polarized groups.
    % Evaluation results on real graphs: both \ours and \focg find multiple non-overlapping polarized groups.
    % The figure shows the distribution of the 3 evaluation metrics. 
    For \sbr, smaller values indicates better performance, while for \ham and \pc, larger values are better.
  }
  \label{fig:realgraphs}
\end{figure*}

We make the following observations:

\spara{Effect of correlation parameter $\kappa$.}
As a part of the answer to \textbf{Q1},
Figures~\ref{fig:syn}(a) and (d) indicate that 
smaller $\kappa$ produces solutions with larger $\vol(\Sone \cup \Stwo)$.
This phenomenon can be explained by Claim~\ref{claim:relaxation}.
Recall that $\kappa=\sqrt{1/k}$, where $k$ upper bounds the solution volume in \localpoldiscreteshort and
$\kappa$ is the correlation parameter in \localpols. 
A smaller $\kappa$ implies a larger $k$.
Thus, the constraint $\vol(\Sone \cup \Stwo) \le k \vol(\seedp \cup \seedn)$ in \localpoldiscreteshort encourages solutions with larger $\vol(\Sone \cup \Stwo)$ if a larger $k$ is given, 
% As an implication from this observation, if small solutions are more desirable in practice, $\kappa$ should be set large. 

With respect to the second part of the question \textbf{Q1}, 
we observe that larger values of $\kappa$ make \ours more noise-resistant.
Figure~\ref{fig:syn}(c) shows directly that $\kappa=0.9$ achieves the best \ap scores.
Internally, large $\kappa$ values make \ours select solutions with small volume, 
which increases \ap scores.
This concludes the answer to \textbf{Q1}.

% \spara{Answering \textbf{Q2}}
\spara{Effect of noise.}
In Figures~\ref{fig:syn}(a), (b), and (c),
we vary $\noise$ from $0.01$ to $0.3$
and set seed size equal to 2, i.e., $\abs{\seedp} = \abs{\seedn} = 1$).
% We wish to find the answer to \textbf{Q2} from Figure~\ref{fig:syn}(a) and (b).
Thus, with respect to our research question \textbf{Q2}, 
we observe that as we increase $\noise$,
\ours is more likely to pick some $\Sone, \Stwo$ other than $\Sonetrue, \Stwotrue$, 
thus, lowering the \ap scores.
We explain this phenomenon next.

First, \ours misses $\Sonetrue, \Stwotrue$ when there is another $\Sone, \Stwo$ that gives better \sbr, i.e., $\sbr(\Sone, \Stwo) < \sbr(\Sonetrue, \Stwotrue)$. 
This happens when $\noise$ is large enough (shown in Figure~\ref{fig:syn}(b)).
On one hand, as we add more noise,
$\Sonetrue, \Stwotrue$ becomes less polarized i.e., $\sbr(\Sonetrue, \Stwotrue)$ increases.
When there is a sufficiently large amount of noise, 
the polarization structure is destroyed, i.e., $\sbr(\Sonetrue, \Stwotrue) \approx 1$. 
On the other hand, as $\sbr(\Sonetrue, \Stwotrue)$ approaches $1$,
solutions with larger $\vol(\Sone, \Stwo)$ are preferred as it makes $\sbr(\Sone, \Stwo)$ smaller.
This is shown by the increased solution volume in Figure~\ref{fig:syn}(a).
Combining the above observations, it follows that when the noise increases,
the accuracy of \ours finding $\Sonetrue, \Stwotrue$ decreases.
% larger-volume solutions are picked intead of 
% since $\sbr(\Sone, \Stwo) < \sbr(\Sonetrue, \Stwotrue)$.
% This explains the trend in Figure~\ref{fig:syn}(b).

% Coupled with the effect of $\kappa$ on $\vol(\Sone, \Stwo)$,
% we argue the following:

\spara{Effect of seed size.}
In Figures~\ref{fig:syn}(d), (e), and (f), we vary the seed size
on $\Sonetrue$ and $\Stwotrue$ 
from 2 to 20
and we set edge noise $\noise=0.05$.

We notice that the effects on all 3 metrics are limited when $\kappa$ is either too large 
(e.g., $0.9$) or too small (e.g., $0.1$).
In contrast, when $\kappa=0.5$, seed size starts to play a remarkable role.
For example, as $\abs{\seedp \cup \seedn}$ (which is positively correlated with $\vol(\seedp\cup\seedn)$) increases, the solution volume tends to increase (Figure~\ref{fig:syn}(d)).
This can be explained by using an argument similar to the effect of $\kappa$ on solution volume, i.e.,
when $k$ is fixed, % (as $\kappa$ is fixed), 
larger $\vol(\seedp \cup \seedn)$ \textit{allows} $\vol(\Sone \cup \Stwo)$ to increase.
As a consequence, this brings down the ratio $\sbr(\Sone, \Stwo) / \sbr(\Sonetrue, \Stwotrue)$ 
(Figure~\ref{fig:syn}(e)), which further decreases \ap scores (Figure~\ref{fig:syn}(f)).

So with respect to question \textbf{Q3}, 
we conclude that adding more seeds does not necessarily improve the \ap score.
The value of $\kappa$ affects how much influence the seed number can exert.
The influence is limited when $\kappa$ is either too large or too small. 
On the other hand, when $\kappa$ lies in between, more seeds encourage \ours to produce larger-volume solutions, reducing \ap scores. 
 % including more nodes not in $\Sonetrue, \Stwotrue$.

% \cnote{Aris: this is a bit counter intuitive as the objective is the bipartiteness ratio $\beta$
%   and $\vol(\Sone \cup \Stwo) \le k \vol(\seedp \cup \seedn)$ is just a constraint, i.e.,
%   $\vol(\Sone \cup \Stwo)$ does not really have to be ``pushed up''.}

% \cnote{Han: edited a bit, please check.}

\subsection{Evaluation on real-world graphs}
\label{subsec:comparative}

Next, we compare \ours with \focg and evaluate their capability to finding high-quality polarized communities in real-world graphs.
Since \focg is ``query-less'' and tries to enumerate polarized communities,
we make the following adaptation to make them comparable.
We randomly seed \ours over multiple rounds and collect the results.\footnote{We 
use a simple heuristic to avoid enumerating all possible seeds.
Nodes $u$ and $v$ are considered seeds for $\Sone$ and $\Stwo$ respectively if
$(u, v) \in \negedges$ and $\posdegree(u) \ge t$ and $\posdegree(v) \ge t$,
where $t$ is some pre-defined positive number.}
%
% \footnote{\label{seed-details}Details for setting $t$ and the filtering process are given in the Appendix.}
As \ours can produce communities that overlap,
for fairness, we keep only one of the overlapping communities
\footnote{
  We scan the communities sequentially (in random order), keep track of the nodes that are covered so far and
  drop any communities that intersect with the covered nodes. 
}.
We set $\kappa=0.9$ as we observe that this choice is effective on a wide range of real-world graphs. 
%  and it is not necessary to do exhaustive search to obtain benefits.
In addition, $\kappa=0.9$ ensures that the seed nodes have dominant magnitude in the solution, 
i.e., $\abs{\vxopt_s}$ dominates at $s \in \seedp\cup\seedn$. 

% \cnote{Aris: Which of the overlapping communities you keep? Random? Best?
% Also, deciding which community to keep may affect other communities.
% }

% \cnote{Han: explained the process in footnote, please check.}

% \footnoteref{seed-details}

\spara{Evaluation metrics.} We use 3 metrics to measure quality:
\squishlist
\item[(a)]
    \textbf{Signed bipartiteness ratio $\beta$}, the objective of our problem;
\item[(b)]
    \textbf{\ham}, the harmonic mean of $\coh$ and $\opp$ measures, defined by Chu et al.~\cite{chu2016finding}. In particular,
    \begin{align*}
      \coh(\Sone, \Stwo) &= \frac{1}{2} [d^{\expplus}(\Sone) + d^{\expplus}(\Stwo)] \\
      \opp(\Sone, \Stwo) &= d^{\expminus}(\Sone, \Stwo)
    \end{align*}
    with $d^{\expplus}(C)=\frac{2\abs{\posedges(C)}}{\abs{C}(\abs{C}-1)}$ and $d^{\expminus}(\Sone, \Stwo)=\frac{\abs{\negedges(\Sone, \Stwo)}}{\abs{\Sone} \abs{\Stwo}}$;
  \item[(c)] {\textbf{\pc}~\cite{bonchi2019discovering}, which counts the number of edges that agree with the polarized structure and penalizes large communities:
      \[
        \pc(\Sone, \Stwo) = \frac{\abs{\posedges(\Sone) \cup \posedges(\Stwo)} + 2\abs{\negedges(\Sone, \Stwo)}}{\abs{\Sone \cup \Stwo}}.
      \]
    % In matrix form:
    % $\pc(\Sone, \Stwo)=\vx^T \adjm \vx / \vx^T \vx$
    % where $\vx(u) = 1$ if $u \in \Sone$, $\vx(u) = -1$ if $u \in \Stwo$ and 0 othewise. 
  }
\squishend

\spara{Results.} 
The distributions of the 3 evaluation metrics for both methods are shown in Figure~\ref{fig:realgraphs}.
In Figure~\ref{fig:realgraphs}(a), \ours always outperforms \focg w.r.t. the median of $\sbr$,
which is expected because \ours optimizes $\sbr$. 
In terms of \ham (Figure~\ref{fig:realgraphs}(b)), our method achieves better median value on 4 out of the 5 graphs (except \dswiki).
% The advantage is especially obvious on \dsbc and \dsep. 
This result is somehow surprising because \focg optimizes an objective closer to \ham than our approach. 
Last in Figure~\ref{fig:realgraphs}(c), there is no clear winner for the \pc metric.
% However, in terms of 75th percentile, \pc outperforms slightly on 4 out of 5 graphs (except \dsref). 

\begin{figure*}[h]
  \centering
  \begin{tabular}{ccc}
    % \addfigure{{fair-cheating}}
    \resizebox{.25\linewidth}{!}{
    % This file was created by matplotlib2tikz v0.7.4.
\begin{tikzpicture}

\definecolor{color1}{rgb}{1,0,1}
\definecolor{color0}{rgb}{0,1,1}

\begin{axis}[
tick align=outside,
x grid style={white!69.01960784313725!black},
xmajorticks=false,
xmin=-0.910565442081035, xmax=1.10089160011798,
xtick style={color=black},
y grid style={white!69.01960784313725!black},
ymajorticks=false,
ymin=-0.832779812512302, ymax=0.777800355956462,
ytick style={color=black},
axis lines=none
]

\tikzset{%
  line style/.style={%
    line width=1.5pt
  }
}

\path [line style, draw=red, draw opacity=0.5, dash pattern=on 3.7pt off 1.6pt]
(axis cs:0.69447602268552,0.4852424080637)
--(axis cs:-0.412532782554274,-0.48777040478613);

\path [line style, draw=red, draw opacity=0.5, dash pattern=on 3.7pt off 1.6pt]
(axis cs:0.82724544531964,0.600378299685953)
--(axis cs:-0.412532782554274,-0.48777040478613);

\path [line style, draw=red, draw opacity=0.5, dash pattern=on 3.7pt off 1.6pt]
(axis cs:0.473611909251009,0.56570070285811)
--(axis cs:-0.619930977657831,-0.363821502231698);

\path [line style, draw=red, draw opacity=0.5, dash pattern=on 3.7pt off 1.6pt]
(axis cs:0.473611909251009,0.56570070285811)
--(axis cs:-0.412532782554274,-0.48777040478613);

\path [line style, draw=red, draw opacity=0.5, dash pattern=on 3.7pt off 1.6pt]
(axis cs:0.387054302430582,0.690577058369828)
--(axis cs:-0.619930977657831,-0.363821502231698);

\path [line style, draw=red, draw opacity=0.5, dash pattern=on 3.7pt off 1.6pt]
(axis cs:-0.619930977657831,-0.363821502231698)
--(axis cs:0.277851228575729,0.61679012842704);

\path [line style, draw=red, draw opacity=0.5, dash pattern=on 3.7pt off 1.6pt]
(axis cs:0.277851228575729,0.61679012842704)
--(axis cs:-0.32362050350924,-0.645720725945381);

\path [line style, draw=red, draw opacity=0.5, dash pattern=on 3.7pt off 1.6pt]
(axis cs:0.277851228575729,0.61679012842704)
--(axis cs:-0.209516730487974,-0.738327152550078);

\path [line style, draw=red, draw opacity=0.5, dash pattern=on 3.7pt off 1.6pt]
(axis cs:0.277851228575729,0.61679012842704)
--(axis cs:-0.412532782554274,-0.48777040478613);

\path [line style, draw=red, draw opacity=0.5, dash pattern=on 3.7pt off 1.6pt]
(axis cs:1,0.263984485960453)
--(axis cs:-0.412532782554274,-0.48777040478613);

\path [line style, draw=red, draw opacity=0.5, dash pattern=on 3.7pt off 1.6pt]
(axis cs:0.844033859309476,0.351299024267245)
--(axis cs:-0.412532782554274,-0.48777040478613);

\path [line style, draw=blue, draw opacity=0.5]
(axis cs:-0.246189391207493,-0.598601092911818)
--(axis cs:-0.32362050350924,-0.645720725945381);

\path [line style, draw=blue, draw opacity=0.5]
(axis cs:-0.246189391207493,-0.598601092911818)
--(axis cs:-0.209516730487974,-0.738327152550078);

\path [line style, draw=blue, draw opacity=0.5]
(axis cs:-0.246189391207493,-0.598601092911818)
--(axis cs:-0.412532782554274,-0.48777040478613);

\path [line style, draw=blue, draw opacity=0.5]
(axis cs:-0.798096442006021,-0.435700982528548)
--(axis cs:-0.619930977657831,-0.363821502231698);

\path [line style, draw=blue, draw opacity=0.5]
(axis cs:-0.798096442006021,-0.435700982528548)
--(axis cs:-0.80533809246872,-0.331409108216697);

\path [line style, draw=blue, draw opacity=0.5]
(axis cs:-0.80533809246872,-0.331409108216697)
--(axis cs:-0.619930977657831,-0.363821502231698);

\path [line style, draw=blue, draw opacity=0.5]
(axis cs:0.69447602268552,0.4852424080637)
--(axis cs:0.82724544531964,0.600378299685953);

\path [line style, draw=blue, draw opacity=0.5]
(axis cs:0.69447602268552,0.4852424080637)
--(axis cs:0.844033859309476,0.351299024267245);

\path [line style, draw=blue, draw opacity=0.5]
(axis cs:0.69447602268552,0.4852424080637)
--(axis cs:0.473611909251009,0.56570070285811);

\path [line style, draw=blue, draw opacity=0.5]
(axis cs:-0.497500523329757,-0.389695285011918)
--(axis cs:-0.619930977657831,-0.363821502231698);

\path [line style, draw=blue, draw opacity=0.5]
(axis cs:-0.497500523329757,-0.389695285011918)
--(axis cs:-0.412532782554274,-0.48777040478613);

\path [line style, draw=blue, draw opacity=0.5]
(axis cs:0.0950573166987782,0.598157364564689)
--(axis cs:0.277851228575729,0.61679012842704);

\path [line style, draw=blue, draw opacity=0.5]
(axis cs:0.473611909251009,0.56570070285811)
--(axis cs:0.387054302430582,0.690577058369828);

\path [line style, draw=blue, draw opacity=0.5]
(axis cs:0.473611909251009,0.56570070285811)
--(axis cs:0.277851228575729,0.61679012842704);

\path [line style, draw=blue, draw opacity=0.5]
(axis cs:0.387054302430582,0.690577058369828)
--(axis cs:0.277851228575729,0.61679012842704);

\path [line style, draw=blue, draw opacity=0.5]
(axis cs:-0.619930977657831,-0.363821502231698)
--(axis cs:-0.686604641049421,-0.181083218014751);

\path [line style, draw=blue, draw opacity=0.5]
(axis cs:-0.619930977657831,-0.363821502231698)
--(axis cs:-0.412532782554274,-0.48777040478613);

\path [line style, draw=blue, draw opacity=0.5]
(axis cs:-0.209516730487974,-0.738327152550078)
--(axis cs:-0.32362050350924,-0.645720725945381);

\path [line style, draw=blue, draw opacity=0.5]
(axis cs:1,0.263984485960453)
--(axis cs:0.844033859309476,0.351299024267245);

\path [line style, draw=blue, draw opacity=0.5]
(axis cs:-0.412532782554274,-0.48777040478613)
--(axis cs:-0.32362050350924,-0.645720725945381);

\draw[draw=color0,fill=color0,opacity=0.5] (axis cs:-0.8,0.6) circle (0.04);
\draw[draw=color1,fill=color1,opacity=0.5] (axis cs:-0.8,0.4) circle (0.04);
% \addplot [only marks, draw=color0, fill=color0, opacity=0.5, colormap/viridis]
\addplot [only marks, draw=color0, fill=color0, opacity=0.5]
table{%
x                      y
-0.686604641049421 -0.181083218014751
-0.619930977657831 -0.363821502231698
-0.209516730487974 -0.738327152550078
-0.798096442006021 -0.435700982528548
-0.80533809246872 -0.331409108216697
-0.32362050350924 -0.645720725945381
-0.246189391207493 -0.598601092911818
-0.412532782554274 -0.48777040478613
-0.497500523329757 -0.389695285011918
};
% \addplot [only marks, draw=color1, fill=color1, opacity=0.5, colormap/viridis]
\addplot [only marks, draw=color1, fill=color1, opacity=0.5]
table{%
x                      y
0.277851228575729 0.61679012842704
1 0.263984485960453
0.69447602268552 0.4852424080637
0.387054302430582 0.690577058369828
0.82724544531964 0.600378299685953
0.473611909251009 0.56570070285811
0.844033859309476 0.351299024267245
0.0950573166987782 0.598157364564689
};
% \addplot [only marks, draw=green!50.19607843137255!black, fill=green!50.19607843137255!black, colormap/viridis]
% table{%
% x                      y
% -0.619930977657831 -0.363821502231698
% -0.412532782554274 -0.48777040478613
% };
% \addplot [draw=none, draw=blue, fill=blue, colormap/viridis]
% table{%
% x                      y
% -0.5 -0.5
% 0.5 -0.5
% 0.5 0.5
% -0.5 0.5
% -0.5 -0.5
% };
\node at (axis cs:-0.51189391207493,-0.658601092911818)[
  scale=1.3,
  text=black,
  rotate=0.0
]{trusty};
\node at (axis cs:-0.414930977657831,-0.273821502231698)[
  scale=1.3,
  text=black,
  rotate=0.0
]{fair};
\node at (axis cs:0.104516730487974,-0.748327152550078)[
  scale=1.3,
  text=black,
  rotate=0.0
]{faithful};
\node at (axis cs:-0.681604641049421,-0.141083218014751)[
  scale=1.3,
  text=black,
  rotate=0.0
]{blonde};
\node at (axis cs:-0.507532782554274,-0.447770404786131)[
  scale=1.3,
  text=black,
  rotate=0.0
]{honest};
\node at (axis cs:0.292054302430582,0.730577058369828)[
  scale=1.3,
  text=black,
  rotate=0.0
]{foul};
\node at (axis cs:0.16851228575729,0.50679012842704)[
  scale=1.3,
  text=black,
  rotate=0.0
]{cheating};
\node at (axis cs:0.73224544531964,0.640378299685953)[
  scale=1.3,
  text=black,
  rotate=0.0
]{lying};
\node at (axis cs:0.855,0.153984485960453)[
  scale=1.3,
  text=black,
  rotate=0.0
]{designing};
\node at (axis cs:0.749033859309476,0.391299024267245)[
  scale=1.3,
  text=black,
  rotate=0.0
]{sly};
\node at (axis cs:-0.7,0.58)[
  scale=1.4,
  anchor=base west,
  text=black,
  rotate=0.0
]{ $C_1$};
\node at (axis cs:-0.7,0.36)[
  scale=1.4,
  anchor=base west,
  text=black,
  rotate=0.0
]{$C_2$};
% \node at (axis cs:-0.7,0.25)[
%   scale=0.9,
%   anchor=base west,
%   text=black,
%   rotate=0.0
% ]{seeds in C1};
% \node at (axis cs:-0.7,0.1)[
%   scale=0.9,
%   anchor=base west,
%   text=black,
%   rotate=0.0
% ]{seeds in C2};
\end{axis}

\end{tikzpicture}
    }
    &
      \resizebox{.25\linewidth}{!}{
      \input{data/casestudy3}
    }
    & \includegraphics[width=0.28\linewidth]{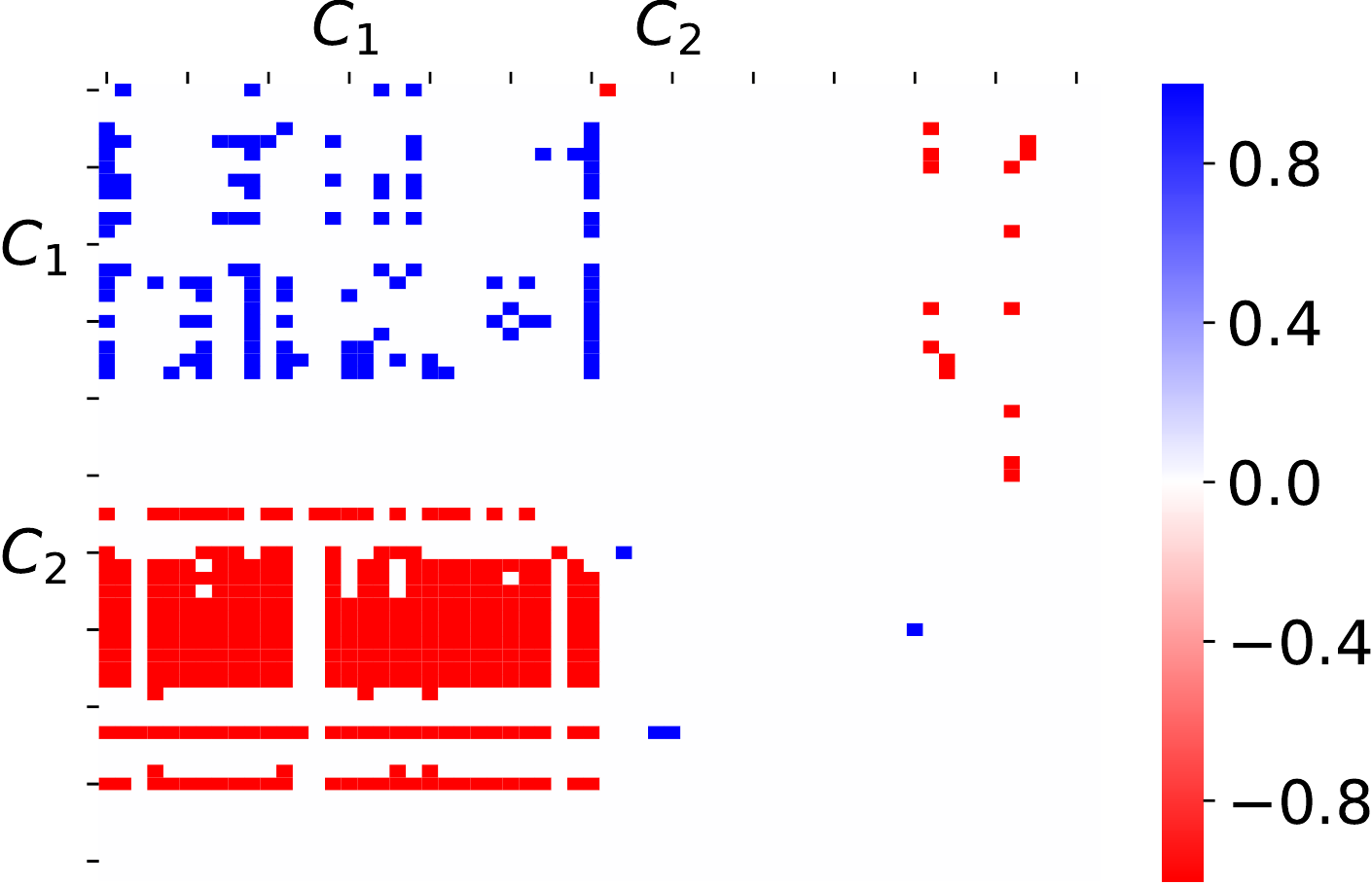} \\
    (a) \textit{fair} as \textit{without cheating}   & (b) \textit{fair} as \textit{not excessive}  & (c) unilateral distrust in \dsbc. \\
    e.g., a \textit{fair} game & e.g., \textit{fair} amount of time &  adjacency matrix of $(\Sone, \Stwo)$ is shown
  \end{tabular}
  \caption{
    (a) and (b): overlapping communities centered on \textit{fair} in the \dsword graph. 
    % \newline
    Seed nodes are: 
    (a) $\seedp=\set{\text{\textit{fair}, \textit{honest}}}, \seedn=\set{\text{\textit{cheating}}}$; 
    (b) $\seedp=\set{\text{\textit{fair}, \textit{modest}}}, \seedn=\set{\text{\textit{extreme}}}$;
    % \newline
    (c) the adjacency matrix of one polarized community in \dsbc graph.
    It highlights a cohesive group ($C_1$) that ``distrusts'' a non-cohesive group ($C_2$).
    The seed set is one node in $C_1$. Positive (resp. negative) edges are represented as blue (resp. red) pixels.
  }
  \label{fig:casestudy}
\end{figure*}

\subsection{Case studies on real-world graphs}
\label{subsec:casestudy}
% Last, we concretise the usefulness of our method via two examples.

\spara{Overlapping communities in \dsword.}
% Overlapping communities exist in a variety of unsigned networks~\cite{yang2014overlapping}.
% We give evidence for the same fact on signed graphs and demonstrate our method's usefulness in finding them in \dsword graph.
We demonstrate the usefulness of our method in finding overlapping communities on the \dsword graph. 
Inspired by the phenomenon of polysemy in natural language, 
we consider \textit{fair} and two of its meanings, \textit{without cheating} and \textit{not excessive}.
For each meaning, we seed our method with words of both similar and opposite meaning. 
The goal is to find antonyms and synonyms of \textit{fair} of the given meaning.
The result is shown in Figures~\ref{fig:casestudy}(a) and (b).
% \footnote{The inclusion of \textit{blonde} in Figure~\ref{fig:casestudy}a illustrates the ``free-rider'' effect~\cite{wu2015robust}}. 

\spara{Unilateral distrust in \dsbc.}
We highlight the capability of our method in finding subgraph pairs that have distrust only in one direction. 
We seed \ours only with nodes from one group, 
e.g., $\seedp \neq \emptyset$ and $\seedn = \emptyset$.
We found one community (of size $62$) that has a cohesive band $\Sone$, which radiates out many negative edges
towards another band $\Stwo$, which is not cohesive at all.
We visualized the adjacency matrix\footnote{
  For visualization, we use the original adjacency matrix here, which is asymmetric.
}
of this community in Figure~\ref{fig:casestudy}(c). 
Note that since \localpols does not enforce connectivity on neither $\Sone$ nor $\Stwo$,
it is understandable that $\Stwo$ is not well connected within itself.

% \cnote{Aris: it is not very clear what is happening here.
%   Can we expand the explanation?}
% \cnote{Han: added more, please check.}

\subsection{Running time on \dswikilarge.}
\label{subsec:scalability}

To study the scalability of our method, 
we produce a large graph \dswikilarge by 
($i$) adding dummy nodes into \dswiki so that $\abs{\nodes}=\text{1 million}$ and
($ii$) adding edges so that both the average degree and the ratio of negative edges remain the same with respect to \dswiki.
% $(ii)$ creating edges from dummy nodes so that each node has average degree of the original \dswiki
% and the number of negative edges is determined so that ratio of negative edges remains the same as \dswiki.
Then, we randomly select 100 seeds using the same heuristic described in Section~\ref{subsec:comparative} and report the average running time and its standard deviation.
We decompose running time into:
(1) approximating $\xopt$ takes {17.2 ($\pm$ 8.1) seconds} and
(2) rounding $\xopt$ to obtain$(\Sone, \Stwo)$  takes {1.4 ($\pm$ 0.1) seconds}.
We conclude that our method scales nicely
and can be applied to find local polarized communities in large signed graphs.

\section{Conclusion}
\label{sec:conclusion}
We propose a local spectral method to find polarized communities in signed graphs.
We characterize the solutions in close-form using techniques in duality theory and linear algebra.
Based on results from spectral graph theory on signed graphs, our method yields solutions with approximation guarantees with respect to a local variant of signed Cheeger's constant.
We empirically demonstrate our method's scalability and effectiveness in finding polarized communities on synthetic and real-world graphs. 

Our works opens interesting questions for future research. 
For example,
$(i)$ can we extend the current work to finding $k$-way polarized communities ($k>2$)?
$(ii)$ how does the signed bipartiteness ratio compare with other polarization measures, such as \textit{Polarity} proposed in ~\cite{bonchi2019discovering}?
$(iii)$ how to enforce graph connectivity on both bands in a polarized community?

\cnote{Aris: Directions for future work?}
\cnote{Han: check above.}

\spara{Acknowledgments.}
This work was supported by 
three Academy of Finland projects (286211, 313927, 317085), 
the EC H2020RIA project ``SoBigData++'' (871042), 
and
the Wallenberg AI, Autonomous Systems and Software Program (WASP) funded by Knut and Alice Wallenberg Foundation.

\appendix
\section{Appendix}
\subsection{Proof of Claim~\ref{claim:rank1}}

We first present a few lemmas which will be used to prove Claim~\ref{claim:rank1}.

\begin{lemma}
  \label{lm:rankn}
  If $\eigval>0$ and $\eigval > \alpha$, the rank of $\lap - \alpha \dmat$ is $n$.
\end{lemma}
\begin{proof}
  Our assumption that $\graph$ is unbalanced (Theorem~\ref{thm:solution}) translates into $\eigval>0$~\cite{kunegis2010spectral}. 
  We rewrite $\lap - \alpha \dmat = \dmat^{1/2}(\nlap - \alpha I)\dmat^{1/2}$.
  We slightly abuse the notation and denote $\lambda_i(M)$ as the $i$th smallest eigenvalue of $M$. 
  Then we have $\lambda_i(\nlap - \alpha I) = \lambda_i(\nlap) - \alpha$ for all $i=1, \ldots, n$.
  As $\eigval > \alpha$,  it holds that $\lambda_i(\nlap) - \alpha > 0$ for all $i$.
  Given $n$ positive eigenvalues, the rank of $\nlap - \alpha I$ is $n$.
  Multiplying $\nlap - \alpha I$ by positive definite matrix $\dmat^{1/2}$ does not change its rank.% , so the rank of $\lap - \alpha \dmat$ is $n$.
\end{proof}

% We next present a few linear algebra results.

\begin{lemma}
  \label{lm:nullspace}
  Given $A, B \in \real^{m\times m}$ and $A, B \geneq 0$, $A \mprod B = 0$ implies $\mrange(A) \subseteq \mnull(B)$, where $\mrange(A)$ denotes the range of $A$ and $\mnull(B)$ denotes null space of $B$.
\end{lemma}
\begin{proof}
  It is $A \mprod B = B \mprod A = \trace(B^T A) = \trace(BA) = \sum_{i=1}^m \lambda_i(BA) = 0$.
  $A, B \geneq 0$ implies $\lambda_i(BA) \ge 0$ for all $i$. 
  Combining the above, we have $\lambda_i(BA)=0$ for all $i=1,\ldots,m$.
  
  Then $BA$ can be diagonalized into $S \Lambda S^{-1}$, where $S$ has eigenvectors of $BA$ as its columns and $\Lambda$ is a diagonal matrix with $\lambda_i(BA)=0$ as the diagonal entries.
  In other words, for any $y \in \real^{m}$, $BAy=0$ always holds, note that $Ay$ corresponds to $\mrange(A)$.
  % {\color{blue} I think this second paragraph is superfluous, considering $\|BAy\|\leq\|BA\|\|y\|\leq 0$.}
\end{proof}

\begin{lemma}
  \label{lm:rank-reduction}
  $-\beta^*(\dmat s)(\dmat s)^T$ reduces the rank of $L_G-\alpha^*L_{K_n}$ by 1.
\end{lemma}
\begin{proof}
  Denote $B=(\dmat s)(\dmat s)^T$, $A=L_G-\alpha^*L_{K_n}$.
  Both $B$ and $A$ are positive semi-definite, and therefore their eigenvalues are non-negative.
  We can consider the eigenvalue decomposition $B=Q\mathcal{D}Q^T$. Since $B$ has rank equal to 1, $\mathcal{D}$ is a diagonal matrix with a single non-zero element.
  The spectrum of $Q^TAQ$ and $A$ is the same, as this is a similarity transformation. Therefore, we can analyze the difference $Q^T(A-\beta^*B)Q=Q^TAQ-\beta^*\mathcal{D}$.

  Now, observe that $\sum_i\lambda_i(A)=\trace(A)=\trace(Q^TAQ)$, where $\lambda_i(A)$ is the $i$-th eigenvalue of $A$. Since $Q^TAQ-\beta^*D$ results in the modification of a single element of the diagonal of $Q^TAQ$, by increasing $\beta^*$ we can make $tr(Q^TAQ)$ arbitrarily close to $-\infty$. As the eigenvalues of a matrix are continuous functions of its entries, we have that for some $\beta^*$, at least one eigenvalue of $Q^TAQ-\beta^*\mathcal{D}$ must become zero.

  The fact that the difference in rank is at most 1 follows trivially from the fact that only one row is modified.
\end{proof}

\renewcommand*{\proofname}{\textbf{Proof of Claim~\ref{claim:rank1}}}
\begin{proof}

  Multiplying \evecT and \evec on both sides of Eq.~\ref{eq:feas2}, we get

  \vspace{-0.4cm}
  
  \begin{align*}
    & \evecT \lap \evec - \alphaopt \evecT \dmat \evec - \betaopt \evecT  (\dmat \vs)(\dmat \vs)^T \evec  \\
    =\:&\eigval - \alphaopt - \betaopt (\evecT \dmat \vs)^2 \ge\: 0 
  \end{align*}

  The second line follows by $\inprodD{\evec}{\evec} = 1 \text{ and }  \evecT \lap \evec = \eigval$.
  As $\betaopt (\evecT \dmat \vs)^2 \ge 0$, it must be the case $\eigval \ge \alphaopt$.

  \noindent\textbf{Case 1:} $\eigval = \alphaopt$.
  As we assume $\vs^T \dmat \evec \neq 0$, $\betaopt = 0$. 
  Plugging this into Eq.~\ref{eq:cs2}, we have

  \vspace{-0.4cm}
  
  \begin{align*}
    \Xopt \mprod (\lap - \alphaopt \dmat) &= 0 \quad \Rightarrow \Xopt \mprod \lap - \alphaopt \Xopt \mprod  \dmat = 0 \\
    \Rightarrow \Xopt \mprod \lap - \alphaopt &= 0 \quad  \Rightarrow \Xopt \mprod \lap  = \eigval  \\
  \end{align*}

  \vspace{-0.6cm}
  
  The 2nd line follows by Eq.~\ref{eq:feas1}.
  Decomposing $\Xopt$ as a the outer product of its eigenvectors gives $\Xopt=\sum_i\sigma_iww^T$.
  As $X$ is PSD and $\lambda_1=\min_xx^TLx$, we conclude $\Xopt = \evec \evecT$, which has rank 1.

  \noindent\textbf{Case 2:} $\eigval > \alphaopt$.
  Rank of matrix $\lap - \alphaopt \dmat$ is $n$ (by Lemma~\ref{lm:rankn}).
  From Eq.~\ref{eq:cs2}, we have that the range of $\Xopt$ lies in the null space of $\lap - \alphaopt \dmat - \betaopt \dsds$ (by Lemma~\ref{lm:nullspace}). 
  As $\dsds$ is a rank-1 matrix and using Eq.~\ref{eq:feas2}, we have the rank of $\lap - \alphaopt \dmat$ decreases by 1 after subtracting $\betaopt \dsds$ (Lemma~\ref{lm:rank-reduction}). 
  Therefore, $\Xopt$ has rank 1. 
\end{proof}

\subsection{Efficient rounding of $\vx$}
\label{appendix:fast-sweep}
In Proposition~\ref{pp:signed-sweep},
we use$t \in \real^{+}$ to round the solution from $\vx$ to $(\Sone(t), \Stwo(t))$, where
$\Sone(t) = \set{u \in \nodes \mid {\vx}(u) \ge t}$ and $\Stwo(t) = \set{u \in \nodes \mid {\vx}(u) \le -t}$\footnote{We omit $\vx$ in certain notation when context is clear. }.
Our goal is to select a value $t$ that minimizes $\sbr(\Sone(t), \Stwo(t))$.
As a convention, when we say ``order'', we refer to descending order by default. 
We describe how to find the optimal $t$ in $\bigO(m+n\log(n))$, where $n\log(n)$ is the sorting cost.  

First, it's easy to see that we only evaluate $t$ values from the set $\domain(\vx) = \set{\abs{\vx_i} \mid i=1,\ldots,n}$ and $\abs{\domain(\vx)} \le n$. 
W.l.o.g., we assume $\avx$ is strictly ordered,
e.g., $\avx_i > \avx_{i+1}$, for all $i=1, \ldots, n-1$. 
In other words, thresholding on $t=\avx_i$ takes the top-$i$ nodes ordered by $\avx$.
We denote these nodes as $\Cavx[i]$.
Similarly, we define the top-$i$ nodes ordered by $\vx$ and $\nvx$ as
$\Cvx[i]$ and $\Cnvx[i]$ respectively. 

Further, we partition $\Cavx[i]$ into $\Cvx[j]$ and $\Cnvx[k]$
where $j =\abs{\set{u \in \Cavx[i] \mid \vx_u \ge 0}}$
and $k =\abs{\set{u \in \Cavx[i] \mid \vx_u < 0}}$
% \footnote{To make notation lighter, we make the values of $j$ and $k$ implicitly determined by $\Cavx[i]$ and $\vx$. }.
\footnote{To make notation lighter, $j$ and $k$'s values are implicitly determined by $i$ and $\vx$. }.
$\Cvx[j]$ and $\Cnvx[k]$ relate to $\Sone(t)$ and $\Stwo(t)$ respectively.
% It immediately follows that $j+k \le i$. 

Therefore,
finding $t^{*} = \argmin_{t \in \domain(\vx)}  \sbr(\Sone(t), \Stwo(t))$ is equivalent as finding a position:
        % &= \argmin_{t \in \domain(\vx)}  \frac{2\abs{\posedges(\Sone(t), \Stwo(t))} +  \abs{\negedges(\Sone(t))} + \abs{\negedges(\Stwo(t))}}{\vol(\Sone(t) \cup \Stwo(t))}\\
        % & \quad\quad\quad\quad   + \frac{\abs{\edges(\Sone(t)\cup\Stwo(t), \nodes\setminus(\Sone(t)\cup\Stwo(t)))}}{\vol(\Sone(t) \cup \Stwo(t))}
% \end{align*}
\begin{align*}
  \label{eq:sweep_formula}
  i^{*} &= \argmin_{i \in \set{1,\ldots,n}}  \frac{2\abs{\posedges(\Cvx[j], \Cnvx[k])} +   \abs{\negedges(\Cvx[j])} + \abs{\negedges(\Cnvx[k])}}{\vol(\Cavx[i])} \\
        & \quad\quad\quad\quad + \frac{\abs{\edges(\Cavx[i], \nodes\setminus \Cavx[i])}}{\vol(\Cavx[i])}
\end{align*}
% where $j$ and $k$ are implicitly determined by the choice $i$.
In other words, $t^{*}=\avx_{i^{*}}$.

To make notation lighter, we make the following short-cut notation which depends on vector $\vx$ and position $i$:
\begin{align*}
  \cut_{\vx}[i]    &= \abs{\edges(\Cvx[i], \nodes\setminus\Cvx[i])} & \posin_\vx[i]     &= \abs{\posedges(\Cvx[i])} \\
  \poscut_{\vx}[i] &= \abs{\posedges(\Cvx[i], \nodes\setminus\Cvx[i])} & \negin_\vx[i]  &= \abs{\negedges(\Cvx[i])} \\
  \negcut_{\vx}[i] &= \abs{\negedges(\Cvx[i], \nodes\setminus\Cvx[i])} & & \\                     
\end{align*}

\vspace{-0.5cm}

Our algorithm evaluates at all possible $i$ and pick the optimal position.
However, it achieves efficiency by re-using intermediate results. 
Specifically, we compute the related terms at position $i$ from those at $1,\ldots,i-1$ which are \textit{already computed}. 

Next, we show how to achieve this in $\bigO(m+n)$.

\textbf{Computing $\vol(\Cavx[\cdot])$. }
We define the \textit{\vx-ordered volume}, $\vol_\vx$, which is affected by node ordering induced by $\vx \in \real^{n}$.
Denote the ordering as $\delta: [n] \rightarrow [n]$. 
Then define

\vspace{-0.2cm}

\[\vol_\vx[i]=\sum_{j=1,\ldots,i} \deg(\delta(i))\]

\vspace{-0.1cm}

% This gives more variants based on the vector being used, e.g, $\vol_{\nvx}[i]$ and $\vol_{\avx}[i]$.
We also consider signed variants of \vx-ordered volume, e.g.,
$\posvol_{\vx}[i]$ % and $\negvol_{\vx}[i]$

, which only count degrees from positive edges. 

It is easy to see the following equivalence for $i=1,\ldots,n$:

\vspace{-0.2cm}

\[
  \vol(\Cavx[i]) = \vol_{\avx}[i] 
\]

Then, the following recursion holds for $i=1, \ldots, n$:

\vspace{-0.1cm}

\[
  \vol_{\avx}[i] = \vol_{\avx}[i-1] + \degree(i)
\]

where $\vol_{\avx}[0]=0$.
In other words, $\vol_{\avx}[\cdot]$ can be computed incrementally in $\bigO(m+n)$. 
In a similar fashion, we can compute
$\posvol_{\vx}$, $\posvol_{\nvx}$, $\negvol_{\vx}$, $\negvol_{\nvx}$ and $\negvol_{\avx}$ in $\bigO(m+n)$.

% \paragraph{Computing $\abs{\edges(\Cavx[\cdot], \nodes\setminus\Cavx[\cdot])}$.}
\spara{Computing $\cut_{\avx}[\cdot]$.}
We will use the following matrices to compute $\poscut_{\avx}[i]$ and $\negcut_{\avx}[i]$. 
$\Lp$ and $\Ln$ are the lower triangular matrix of $\Ap$ and $\An$ respectively. 
Given a vector $\vx \in \real^{n}$ and a matrix $M \in \real^{n \times n}$,
we define $M_\vx$ as the \textit{$\vx$-ordered matrix $M$}. 
$M_\vx$ equals $M$ with both rows and columns permuted by the ordering induced by $\vx$.
Using this notation, we have $\Ap$ and $\An$ permuted by the order in $\vx, \nvx, \text{ or } \avx$.

% We define $\posin_\vx[i]$ (resp. $\negin_\vx[i]$) as the number of positive (resp. negative) edges in the subgraph induced by top-$i$ nodes from the ordering by $\vx$.
It is easy to see the following recursions hold:

\vspace{-0.2cm}

\begin{align*}
  \posin_\avx[i] &= \posin_\avx(i-1) + 2\;\sum_{j=1}^n\Lp_\avx(i, j) \\
  \negin_\avx[i] &= \negin_\avx(i-1) + 2\;\sum_{j=1}^n\Ln_\avx(i, j)
\end{align*}

\vspace{-0.2cm}

Both of which are computable in $\bigO(m+n)$. Since

\vspace{-0.3cm}

\begin{align*}
  \poscut_\avx[i] &= \posvol_\avx[i] - \posin_\avx[i] \\
  \negcut_\avx[i] &= \negvol_\avx[i] - \negin_\avx[i] \\
  \cut_{\avx}[i]  &= \poscut_{\avx}[i] + \negcut_{\avx}[i]
\end{align*}
Therefore, $\cut_\avx[\cdot]$ can be computed in $\bigO(m+n)$. 

% \paragraph{Computing $\abs{\negedges(\Cvx[j])} + \abs{\negedges(\Cnvx[k])}$ given $i$. }
\spara{Computing $\negin_\vx[j] + \negin_\vx[k]$ given $i$. }
% For brevity, we denote $\negin_{\vx}(\cdot) = \abs{\negedges(\Cvx[\cdot])}$
% and similarly, $\negin_{\nvx}(\cdot) = \abs{\negedges(\Cnvx[\cdot])}$. 
% Both $\negin_{\vx}(\cdot)$ and $\negin_{\vx}(\cdot)$ can be computed in $\bigO(m+n)$ as shown previously.
Recall that $j$ and $k$ are determined implicitly by $i$.
% $j =\abs{\set{u \in \Cavx[i] \mid \vx_u \ge 0}}$
% and $k = i-j$ respectively.
In practice, for all $i=1,\ldots,n$, the list of $[j]$ and $[k]$ can be computed in $\bigO(n)$. 
Note that both $\negin_{\vx}(\cdot)$ and $\negin_{\vx}(\cdot)$  are already computed before.
Therefore, this step is done in $\bigO(n)$. 

\spara{Computing $\abs{\posedges(\Cvx[j], \Cnvx[k])}$. }
We observe the following :
\begin{align*}
  &\posedges(\Cvx[j], \Cnvx[k]) \\
  = & \posedges(\Cvx[j], \nodes\setminus\Cvx[j]) \; \cup \; \posedges(\Cnvx[k], \nodes\setminus\Cnvx[k]) \\
  & \setminus \; \posedges(\Cavx[i], \nodes\setminus\Cavx[i])
\end{align*}
It follows immediately:
\[
  \abs{\posedges(\Cvx[j], \Cnvx[k])} = \poscut_\vx[j] + \poscut_{\nvx}[k] - \poscut_\avx[i]
\]
which is computable in $\bigO(m+n)$ for all $i=\enlst{1}{n}$.

\clearpage
\bibliographystyle{ACM-Reference-Format}
\bibliography{references}
\end{document}